\documentclass[twoside, 11pt, journal, onecolumn]{IEEEtran}

\usepackage{amssymb}
\usepackage{amsmath}
\usepackage{amsthm}
\usepackage[dvips]{graphicx}
\usepackage{subfig}
\usepackage{url}
\usepackage{setspace}

\theoremstyle{plain}
\newtheorem{theorem}{Theorem}
\newtheorem{lemma}{Lemma}
\newtheorem{proposition}{Proposition}
\newtheorem{corollary}{Corollary}

\theoremstyle{definition}
\newtheorem{definition}{Definition}
\newtheorem{example}{Example}
\newtheorem{remark}{Remark}

\begin{document}

\title{\Huge Polar Coding for
Parallel Channels{\footnote{This research was supported by the
Israel Science Foundation (grant no. 1070/07), and by the European
Commission in the framework of the FP7 Network of Excellence in
Wireless Communications (NEWCOM++).}}}

\author{\vspace{1cm} \IEEEauthorblockN{Eran Hof\IEEEauthorrefmark{1} \ \ \
Igal Sason\IEEEauthorrefmark{1} \footnote{Igal Sason is the corresponding author (E-mail:
sason@ee.technion.ac.il).} \ \ \ Shlomo
Shamai\IEEEauthorrefmark{1} \ \ \  Chao Tian\IEEEauthorrefmark{2}} \\[0.2cm]
\IEEEauthorblockA{\IEEEauthorrefmark{1} Department of Electrical Engineering\\
Technion -- Israel Institute of Technology\\
Haifa 32000, Israel \\
E-mails: eran.hof@gmail.com,
\{sason@ee, sshlomo@ee\}.technion.ac.il }  \\[0.2cm]
\IEEEauthorblockA{\IEEEauthorrefmark{2}AT\&T Labs-Research
\\ 180 Park Ave. \\
Florham Park, NJ 07932 \\
Email: tian@research.att.com}}

\maketitle


\begin{abstract}
A capacity-achieving polar coding scheme is introduced for reliable
communications over a set of parallel channels. The parallel channels are assumed to
be arbitrarily-permuted memoryless binary-input output-symmetric
(MBIOS) channels. A coding scheme is first provided for the particular case
where the parallel channels form a set of stochastically degraded
channels. Next, the case of polar coding for the more general case where
the parallel channels are not necessarily degraded is also
considered, and two modifications are provided for this case.
\end{abstract}

\IEEEpeerreviewmaketitle
\section{Introduction}

Channel coding over a set of parallel arbitrarily-permuted
channels is studied in~\cite{WillemsGorolhovParallelChannels}. The
information message in such a setting is encoded into a set of
codewords, all with a common block length. These codewords are
transmitted over a set of parallel discrete memoryless channels
(DMC) where the assignment of codewords to channels is arbitrary.
This assignment is known only to the receiver, which decodes the
transmitted message based on the set of received vectors. In cases
where the capacities of all of the parallel channels are achieved
with a common input distribution, it is proved
in~\cite{WillemsGorolhovParallelChannels} that the capacity for
the considered parallel setting equals the sum of the capacities
of the parallel channels. Such parallel channel models are of
concern when analyzing networking applications, OFDM and BICM
systems.

The coding schemes suggested
in~\cite{WillemsGorolhovParallelChannels} for the considered
parallel setting are based on random-coding and joint-typicality
decoding. One of the main contributions
of~\cite{WillemsGorolhovParallelChannels} is the introduction of a
concatenation of rate-matching codes with parallel copies of a fully
random block code. A rate-matching code is a device that encodes a
single message into a set of messages. It is shown
in~\cite{WillemsGorolhovParallelChannels} that under specific
structural conditions on the rate-matching code, such a concatenated
scheme can achieve the capacity of the set of parallel channels.
Moreover, it is shown how to construct these rate-matching codes
from a set of maximum-distance separable (MDS) codes. The decoding
procedure for the concatenated scheme is based on successive
cancellation and joint-typicality (list) decoding.

Polar codes form a class of capacity-achieving block codes
\cite{ArikanPolarCodes}. These codes are shown to achieve the
capacity of a symmetric DMC with a practical encoding and decoding complexity
(in terms of the block length). Encoding of polar codes is defined
based on a recursive approach. This recursion is a key ingredient
both in proving the capacity-achieving properties of polar codes,
and their successive-cancellation decoding procedure. A set of
predetermined and fixed bits are incorporated in the encoding
procedure of polar codes, and it plays a crucial role in the decoding process.

Parallel polar coding schemes are provided in this paper for
communication over binary-input arbitrarily-permuted memoryless
and symmetric parallel-channels. The particular case where the
parallel channels form a set of stochastically degraded channels
is first addressed. A parallel polar coding scheme is first
provided for this particular case. While the provided scheme
achieves the capacity of degraded parallel channels, it is shown
not to achieve capacity for the general case where the channels
are no longer degraded. Finally, two modifications are provided
for the general case when the channels are not necessarily
degraded. Both of the two modification are shown to achieve the capacity for the case at hand.

The main difference between the proposed coding schemes and the
original polar coding scheme in~\cite{ArikanPolarCodes}, is in the
setting of the predetermined and fixed bits which are incorporated
in the encoding and decoding procedures.
In~\cite{ArikanPolarCodes}, for symmetric DMC, these bits may be
chosen arbitrarily; they are fixed and do not depend on the
transmitted message. For the provided scheme, some of the
concerned bits incorporate an algebraic structure and depend on
the transmitted message. Moreover, the determination of these bits
is based on the structural properties of MDS codes, in a manner
which relates to the rate-matching code
in~\cite{WillemsGorolhovParallelChannels}.

This paper is structured as follows:
Section~\ref{section:preliminaries} provides some preliminary
material. The parallel polar coding scheme is introduced and
analyzed in Section~\ref{section:TheProposedScheme} for the
particular case of degraded parallel channels. Two modified
parallel schemes which achieve the capacity for the case of non-degraded channels
are studied in Section~\ref{section:NonDegradedCase}.
Section~\ref{section:conclusions} concludes the paper.

\section{Preliminaries}
\label{section:preliminaries}

\subsection{Arbitrarily Permuted Parallel Channels}
\label{section:communicationModel}

\begin{figure}
\centering
\includegraphics[width=3.5in]{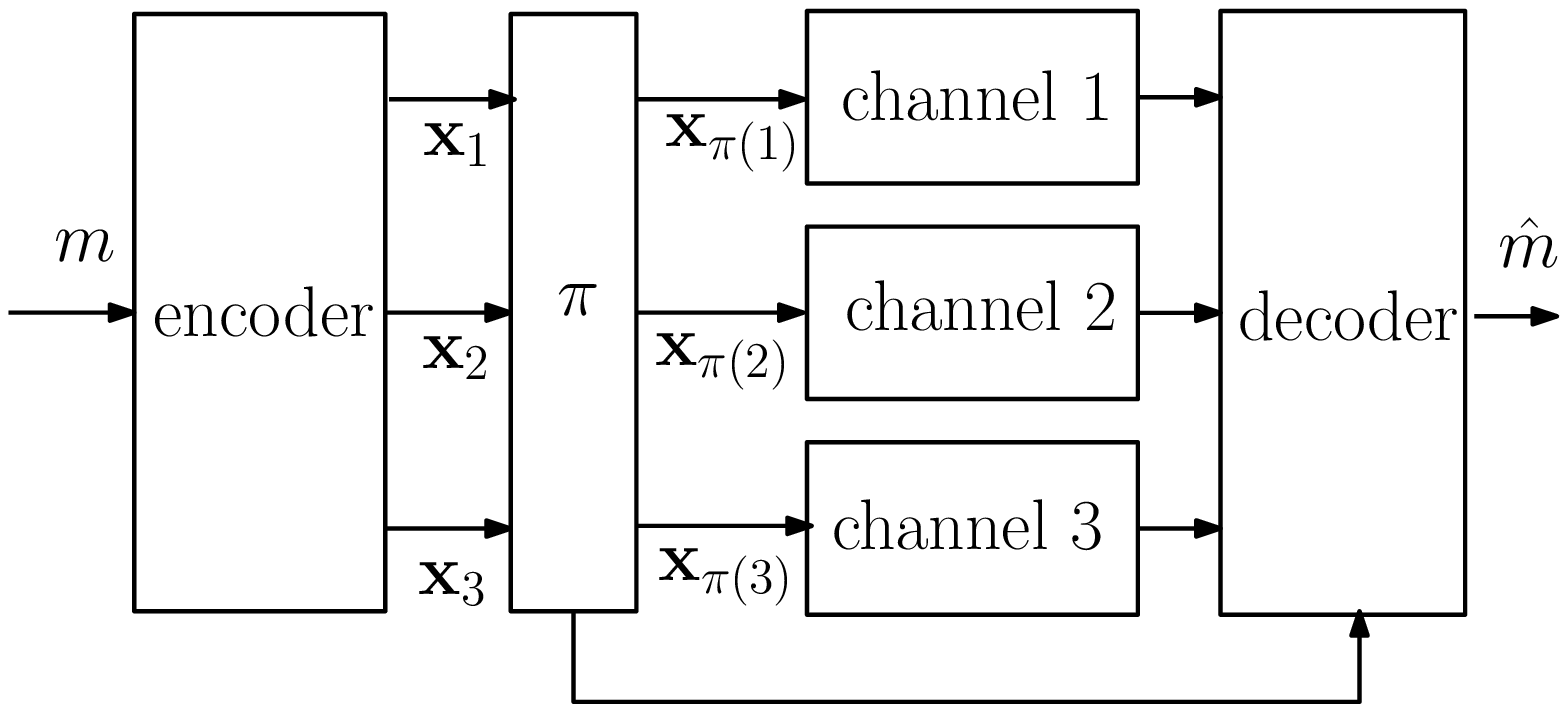}
\caption{\footnotesize{Communication over an arbitrarily-permuted
parallel channel. The particular case of communicating over $S=3$
parallel channels is depicted (taken from~\cite{WillemsGorolhovParallelChannels}).}}
 \label{figure:ArbitrarilyPermutedParallelChannels}
\end{figure}

We consider the communication model in
Figure~\ref{figure:ArbitrarilyPermutedParallelChannels}. A message
$m$ is transmitted over a set of $S$ parallel memoryless channels.
The notation $[S] \triangleq \left\{1,\ldots,S\right\}$ is used in
this paper. All channels are assumed to have a common input
alphabet $\mathcal{X}$, and possibly different output alphabets
$\mathcal{Y}_s$, $s \in [S]$. The transition probability function
of each channel is denoted by $P_{s}(y_s|x)$, where $y_s \in
\mathcal{Y}_s$, $s \in [S]$, and $x \in \mathcal{X}$. For the
particular case depicted in
Figure~\ref{figure:ArbitrarilyPermutedParallelChannels}, the
communication takes place over a set of $S = 3$ parallel channels.
The encoding operation maps the message $m$ into a set of $S$
codewords $\{\mathbf{x}_s \in \mathcal{X}^n\}_{s=1}^S$. Each of
these codewords is of length $n$, and it is transmitted over a
different channel. The mapping of codewords to channels is done by
an arbitrary permutation $\pi:[S]\to[S]$. The permutation $\pi$ is
part of the communication channel model, the encoder has no
control on the arbitrary permutation chosen during the codeword
transmission. The permutation $\pi$ is fixed during the
transmission of the codewords. The set of possible $S$ channels
are known at both the encoder and decoder. The encoder has no
information about the chosen permutation. The decoder, on the
other hand, knows the specific chosen permutation. Formally, the
channel is defined by the following family of transition
probabilities:
\begin{equation*} \label{equation:productChannelDefinition}
\Bigl\{ P\bigl(\mathbf{Y}|\mathbf{X}; \pi\bigr):\ \mathbf{Y}\in
\{\mathcal{Y}_1 \times \mathcal{Y}_2 \times \cdots \times \mathcal{Y}_S \}^n,\
\mathbf{X}\in\mathcal{X}^{s\times n},\ \pi:[S]\to[S]\Bigr\}_{n=1}^\infty
\end{equation*}
where $\mathbf{X} = \left(\mathbf{x}_1,\mathbf{x}_2, \ldots,
\mathbf{x}_S\right)$ are the transmitted codewords, $\mathbf{Y} =
\left(\mathbf{y}_1, \mathbf{y}_2, \ldots, \mathbf{y}_S\right)$ are
the received vectors,
\begin{equation} \label{equation:productChannelNotation}
P\bigl(\mathbf{Y}|\mathbf{X}; \pi\bigr) = \prod_{s=1}^S P_s\bigl(\mathbf{y}_s| \mathbf{x}_{\pi(s)}\bigr)
\end{equation}
is the probability law of the parallel channels, and
$\pi:[S]\to[S]$ is the arbitrarily permutation mapping of
codewords to channels.

The coding problem for this communication model is to
guarantee reliable communication for all possible ($S!$) permutations
$\pi$. This problem is formulated and studied
in~\cite{WillemsGorolhovParallelChannels}.

\begin{definition}[\textbf{Achievable rates and channel capacity}]
\label{definition:AchievableRateCapacity} Consider coded
communication over a set of $S$ arbitrarily permuted parallel
channels. A rate $R>0$ is achievable if there exists a sequence of
encoders and decoders such that for all $\delta >0$ and a
sufficiently large block length $n$
\begin{align}
\label{equation:RateCondition}
& \frac{1}{n} \log_2 M \geq R - \delta\\
\label{equation:ErrorCondition} & P_{\text{e}}^{(\pi)}(n) \leq
\delta, \ \text{for all } S! \text{ permutations } \pi:[S]\to[S]
\end{align}
where $M$ is the number of possible messages and
$P_{\text{e}}^{(\pi)}(n)$ is the average block error probability for
a fixed permutation $\pi$ and block length $n$. The capacity of the
considered model $C_{\Pi}$ is the maximal achievable
rate to satisfy~\eqref{equation:RateCondition}
and~\eqref{equation:ErrorCondition}.
\end{definition}

The following theorem may be derived as a particular case of the
well-known results on the capacity of the compound channel (see,
e.g.,~\cite{LapidothNarayanChannelUncertainty} and reference
therein). Nevertheless, the theorem is stated in the restricted
form as provided in~\cite{WillemsGorolhovParallelChannels}:

\begin{theorem}[\textbf{The capacity of arbitrarily-permutated
memoryless parallel channels \cite{WillemsGorolhovParallelChannels}}]
\label{theorem:capacityOfParallelChannels} {\em Consider the
transmission over a set of $S$ arbitrarily-permutated memoryless
parallel channels. Assume that there is an input distribution that
achieves capacity for all parallel channels. Then, the capacity
$C_{\Pi}$ satisfies
\begin{equation} \label{equation:capacityParallelChannels}
C_{\Pi} = \sum_{s=1}^S C_s
\end{equation}
where $C_s$ is the capacity of the $s$-th channel, $s\in [S]$.}
\end{theorem}


As noted in~\cite{WillemsGorolhovParallelChannels}, if both the encoder and decoder know the actual
permutation $\pi$, then the capacity is clearly given by $\sum_{s=1}^S
C_s$; since in the considered channel model the encoder does not know the actual
permutation, then
\begin{equation*}
C_{\Pi} \leq \sum_{s=1}^S C_s.
\end{equation*}
That equality is achievable is proved in \cite{WillemsGorolhovParallelChannels} using two
different approaches:
\begin{enumerate}
\item A random coding argument and a joint-typicality decoding over product
channels. This coding scheme is based on the notion of product
channels as defined in~\eqref{equation:productChannelNotation}.
Each possible permutation $\pi$ yields a different product
channel. Consequently, there are $S!$ possible product channels. A
properly chosen random code is shown to achieve the capacity
$C_{\Pi}$ under a joint-typicality decoding scheme for all
possible permutations $\pi$.
\item A rate-matching coding scheme that is combined with a random
coding argument, and a sequential joint-typicality decoding. The construction technique for rate-matching codes in~\cite{WillemsGorolhovParallelChannels}, based on MDS codes, provided an important intuition for the parallel polar schemes introduced in the following sections.
\end{enumerate}

\subsection{Polar Codes}
\label{section:polarCodes}

This preliminary section offers a short summary of the basic
definitions and results in \cite{ArikanPolarCodes},
\cite{TelatarArikanPolarRate}, that are essential in the following
sections. For a DMC, polar codes achieve the mutual information
between an equiprobable input and the channel output. It is well
known that the information rate under equiprobable inputs is equal
to the channel capacity of a symmetric DMC.

\begin{definition}[\textbf{Symmetric binary input channels}]
\label{definition:symmetricBDMC} A DMC with a transition probability
$p$, a binary-input alphabet $\mathcal{X} = \{0,1\}$, and an output alphabet
$\mathcal{Y}$ is said to be symmetric if there exists a permutation
$\mathcal{T}$ over $\mathcal{Y}$ such that
\begin{enumerate}
\item The inverse permutation $\mathcal{T}^{-1}$ is equal to $\mathcal{T}$, i.e.,
\begin{equation*}
\mathcal{T}^{-1}(y) = \mathcal{T}(y)
\end{equation*}
for all $y\in\mathcal{Y}$.
\item The transition probability $p$ satisfies
\begin{equation*}
p(y|0) = p(\mathcal{T}(y)|1)
\end{equation*}
for all $y\in\mathcal{Y}$.
\end{enumerate}
\end{definition}


Let $p$ be a transition probability function of a binary-input DMC
with an input-alphabet $\mathcal{X} = \{0,1\}$ and an
output-alphabet $\mathcal{Y}$.
%
Polar codes are defined in~\cite{ArikanPolarCodes} using a
recursive channel synthesizing operation which is referred to as
channel combining. The synthesized channel, after $i\geq 1$
recursive steps has a block input of length $n=2^i$ bits and is
denoted by $p_n$. The output alphabet of the combined channel is
$\mathcal{Y}^n$. The recursive construction of $p_n$ is
equivalently defined by using a linear encoding operation. A
$n\times n$ matrix $G_n$, refereed to as the polar generator
matrix of size $n$, can be recursively defined and the combined
channel can be shown to satisfy:
\begin{equation} \label{equation:equivalentRecursiveConstruction}
p_n(\mathbf{y}|\mathbf{w}) = p(\mathbf{y}|\mathbf{w}G_n)
\end{equation}
for all $\mathbf{y} \in \mathcal{Y}^n$ and $\mathbf{w}\in
\mathbf{X}^n$.

Let $\mathcal{A}_n \subseteq [n]$, and denote by
$\mathcal{A}_n^{\text{c}}$ the complementary set of
$\mathcal{A}_n$, (i.e., $\mathcal{A}_n^{\text{c}} = [n] \setminus
\mathcal{A}_n$). Given a set $\mathcal{A}_n$, a class of coset
codes are formed, all with a code-rate equal to
$\frac{1}{n}|\mathcal{A}_n|$. Over the indices specified by
$\mathcal{A}_n$, the components of $\mathbf{w}$ are set according
to the information bits. The rest of the bits of $\mathbf{w}$ are
predetermined and fixed according to a particular code design. The
set $\mathcal{A}_n$ is referred to as the information set. Polar
codes are constructed by a specific choice of the information set
$\mathcal{A}_n$. This construction can be shown to be equivalent
to a coset code $\mathcal{C}\bigl(G_n\left(\mathcal{A}_n\right),
\mathbf{b} G_n \left(\mathcal{A}_n^{\text{c}}\right)\bigr)$ where
$G_n(\mathcal{A}_n)$ denotes the $|\mathcal{A}_n| \times n$
sub-matrix of $G_n$ defined by the rows of $G_n$ whose indices are
in $\mathcal{A}_n$, $G_n(\mathcal{A}^{\text{c}})$  denotes  the
$|\mathcal{A}_n^{\text{c}}| \times n$ sub-matrix of $G_n$ formed
by the remaining rows in $G_n$, and
\begin{equation} \label{equation:generalCosetCode}
\mathcal{C}(G,\mathbf{c}) \triangleq \left\{ \mathbf{x}:\ \mathbf{x}
= \mathbf{u}G+\mathbf{c}, \ \mathbf{u} \in \mathcal{X}^k\right\}.
\end{equation}

Channel splitting is another important operation that is
introduced in~\cite{ArikanPolarCodes} for polar codes. The split
channels $\{p_n^{(l)}\}_{l=1}^n$, all with a binary input alphabet
$\mathcal{X}$ and output alphabets $\mathcal{Y}^n \times
\mathcal{X}^{l-1}$, $l\in[n]$, are defined according to
\begin{equation} \label{equation:standradPolarChannelSplitting}
p_n^{(l)}(\mathbf{y},\mathbf{w}|x) \triangleq
\frac{1}{2^{n-1}}\sum_{\mathbf{c} \in \mathcal{X}^{n-l}}
p_n\bigl(\mathbf{y}|(\mathbf{w},x,\mathbf{c})\bigr)
\end{equation}
where $\mathbf{y} \in \mathcal{Y}^n$, $\mathbf{w} \in
\mathcal{X}^{l-1}$, and $x \in \mathcal{X}$.
The importance of channel splitting is due to its role in the successive
cancellation decoding procedure that is provided
in~\cite{ArikanPolarCodes}. 
The
decoding procedures iterates over the index $l \in [n]$. If $l\in
\mathcal{A}_n^{\text{c}}$, then the bit $w_l$ is a predetermined and
known bit. Otherwise, the bit $w_l$ is decoded as if it is transmitted over the corresponding split channel $p_n^{(l)}$ in \eqref{equation:standradPolarChannelSplitting}.
%
This decoding
procedure is referred in the following as a standard polar
successive cancellation decoding procedure. It is shown
in~\cite{ArikanPolarCodes} that the successive cancellation decoding
procedure has a complexity of $\mathcal{O}(n \log n)$.

\subsection{Stochastically degraded parallel channels}
\label{subsection:degradedParallelChannels}

The polarization properties of stochastically degraded
parallel-channels are studied in this section.

\begin{definition}[\textbf{Stochastically degraded channels}]
\label{definition:degradedChannels} Consider two memoryless channels
with a common input alphabet $\mathcal{X}$, transition probability
functions $P_1$ and $P_2$, and two output alphabets $\mathcal{Y}_1$
and $\mathcal{Y}_2$, respectively. The channel $P_2$ is a
stochastically degraded version of channel $P_1$ if there exists
a channel $D$ with an input alphabet $\mathcal{Y}_1$ and an output
alphabet $\mathcal{Y}_2$ such that
\begin{equation}\label{equation:channelDegradation}
P_2(y_2|x) = \sum_{y_1 \in \mathcal{Y}_1} P_1(y_1|x) D(y_2|y_1), \ \  \forall x\in\mathcal{X}, y_2\in\mathcal{Y}_2.
\end{equation}
\end{definition}

\begin{lemma}[\textbf{On the degradation of split channels}] \label{lemma:splitChannel4DegradedChannels}
{\em Let $P_1$ and $P_2$ be two transition probability functions
with a common binary input alphabet $\mathcal{X}=\{0,1\}$ and two
output alphabets $\mathcal{Y}_1$ and $\mathcal{Y}_2$,
respectively. For a block length $n$, the split channels of $P_1$
and $P_2$ are denoted by $P_{1,n}^{(l)}$ and $P_{2,n}^{(l)}$,
respectively, for all $l \in [n]$. Assume that the channel $P_2$
is a stochastically degraded version of channel $P_1$. Then, for
every $l\in [n]$, the split channel $P_{2,n}^{(l)}$ is a
stochastically degraded version of the split channel
$P_{1,n}^{(l)}$.}
\end{lemma}

\begin{proof}
The proof follows by induction~\cite{SatishPHD}.
\end{proof}


\begin{definition}[\textbf{Stochastically degraded parallel channels}]
\label{definition:degradedParallelChannels} Let $\{P_{s}\}_{s=1}^S$
be a set of $S$ parallel memoryless channels, and denote the capacity of $P_s$
by $C_s$ for all $s\in[S]$. In addition, assume without loss of
generality that $C_{s}\geq C_{s'}$ for all $1\leq s < s' \leq S$.
The channels $\{P_{s}\}_{s=1}^S$ are stochastically degraded if for
every $1\leq s < s' \leq S$ the channel $P_{s'}$ is a stochastically
degraded version of $P_{s}$.
\end{definition}


\begin{corollary}[\textbf{On monotonic information sets for stochastically degraded parallel channels}]
\label{corollary:informationSetsDegradedChannels} {\em Consider a
set of $S$ memoryless degraded and symmetric parallel channels
$\{P_s\}_{s=1}^S$, with a common binary-input alphabet
$\mathcal{X}$. For every $s\in[S]$, denote the capacity of the
channel $P_s$ by $C_s$, and assume without loss of generality that
\begin{equation*}
C_1 \geq C_2 \geq \cdots \geq C_S.
\end{equation*}
Fix $0 < \beta  \leq \frac{1}{2}$ and a set of rates $\{R_s\}_{s=1}^S$
where
\begin{equation*}
0 \leq R_s \leq C_s, \ \ \forall s\in[S].
\end{equation*}
Then, there exists a sequence of information sets
$\mathcal{A}_n^{(s)}\subseteq[n]$, $s\in[S]$ and $n=2^i$ where $i
\in \mathbb{N}$, satisfying the following properties:
\begin{enumerate}
\item Rate:
\begin{equation} \label{equation:rateProperties}
|\mathcal{A}_n^{(s)}| \geq n R_s, \ \ \forall s \in [S].
\end{equation}
\item Monotonicity:
\begin{equation}
\label{equation:monotonicityProperty4ParallelChannels}
\mathcal{A}_n^{(S)} \subseteq \mathcal{A}_n^{(S-1)} \subseteq \cdots
\subseteq \mathcal{A}_n^{(1)}.
\end{equation}
\item Performance:
\begin{equation} \label{equation:SplitChannelPerformanceProperty}
\Pr\bigl(\mathcal{E}_l(P_s)\bigr) \leq 2^{-n^\beta}
\end{equation}
for all $l \in \mathcal{A}_n^{(s)}$ and $s \in [S]$, and
\begin{align} \label{equation:eventDefinitionErrorInL}
\mathcal{E}_l(p) \triangleq \left\{ p_n^{(l)}(\mathbf{y},
\mathbf{w}^{(l-1)}|w_l) \leq p_n^{(l)}(\mathbf{y},
\mathbf{w}^{(l-1)}|w_l+1)\right\},\ \ l\in[n]
\end{align}
\end{enumerate}}
\end{corollary}

\begin{proof}
The rate and performance properties form immediate consequences of
the polarization properties in~\cite{TelatarArikanPolarRate}.
It is left to prove that the choice of the information set sequences
can be made such that the monotonicity property
in~\eqref{equation:monotonicityProperty4ParallelChannels} is
satisfied. Start with $s=S$. From
~\cite{TelatarArikanPolarRate} it follows that there
exists a sequence of sets $\{\mathcal{A}_n^{(S)}\}$
satisfying~\eqref{equation:rateProperties}
and~\eqref{equation:SplitChannelPerformanceProperty}. Next, fix an
$s'\in[S]$ and assume that for all $s > s'$, the set sequences
$\{\mathcal{A}_n^{(s)}\}$ can be chosen such that the properties
in~\eqref{equation:rateProperties}
and~\eqref{equation:SplitChannelPerformanceProperty} are satisfied,
and in addition
\begin{equation} \label{equation:monotonicityForInduction}
\mathcal{A}_n^{(S)} \subseteq \mathcal{A}_n^{(S-1)} \subseteq \cdots
\subseteq \mathcal{A}_n^{(s'+1)}.
\end{equation}
If $s'=S$ then~\eqref{equation:monotonicityForInduction} is
satisfied in void. The existence of the sequence
$\{\mathcal{A}_n^{(s')}\}$
satisfying~\eqref{equation:rateProperties}
and~\eqref{equation:SplitChannelPerformanceProperty} is already
provided by the polarization properties in~\cite{TelatarArikanPolarRate}.
It is left to verify
that the set sequence can be chosen such that the monotonicity
property
\begin{equation} \label{equation:singleStepMonotonicity}
\mathcal{A}_n^{(s'+1)} \subseteq \mathcal{A}_n^{(s')}
\end{equation}
is kept. Choose an arbitrary index $l \in \mathcal{A}_n^{(s'+1)}$.
It is proved that this index corresponds to the information set for the
channel $P_{s'}$. Specifically, the performance property
in~\eqref{equation:SplitChannelPerformanceProperty} is satisfied for
$s=s'$. Since $P_{s'+1}$ is a degraded version of $P_{s'}$, then
according to Lemma~\ref{lemma:splitChannel4DegradedChannels}, the
split channel $P_{s'+1,n}^{(l)}$ is a degraded version of the split
channel $P_{s',n}^{(l)}$. It is clearly suboptimal to first degrade
the observation vector $\mathbf{y}\in\mathcal{Y}_{s'}$ to create a
vector $\tilde{\mathbf{y}} \in \mathcal{Y}_{s'+1}$, and only then
detect the input bit $x$ for the degraded split channel. However,
the detection error event for the degraded split channel
$P_{s'+1,n}^{(l)}$ satisfies the upper bound
in~\eqref{equation:SplitChannelPerformanceProperty}. As a result,
the optimal detection error for the better split channel
$P_{s',n}^{(l)}$ must also satisfy~\eqref{equation:SplitChannelPerformanceProperty}. Hence, all
the indices in $\mathcal{A}_n^{(s'+1)}$ can be chosen for the set
$\mathcal{A}_n^{(s')}$. The rest of indices are chosen arbitrarily
out of the set of possible indices whose existence is guaranteed by the polarization properties.
The proof follows by induction.
\end{proof}

\begin{remark}{\textbf{On good indices for stochastically degraded channels}}
\label{remark:onIndices4DegradedChannels} In
Corollary~\ref{corollary:informationSetsDegradedChannels}, the
existence of a monotonic sequence of information sets is proved
for a degraded set of channels. A subtle inspection of the proof
shows that the choice of the monotonic sequence of sets can be
carried sequentially. First, the information set of the worst
channel is specified. Then, as is shown
in~\eqref{equation:singleStepMonotonicity}, all the indices that
are ``good'' for the worse channel, are also ``good'' for the
better channel. Here ``good'' is in the sense that the
corresponding Bhattacharyya constants of the split channels (which
form upper bounds on the corresponding decoding error probability)
can be made exponentially low as the block length increases.
Consequently, all that is left to specify are the rest of the
``good'' indices for the better channel (which are ``not good''
for the worse). The construction then follows sequentially.
\end{remark}

\begin{remark} Under the assumptions in Corollary~\ref{corollary:informationSetsDegradedChannels},
the capacity $C_s$ for each of the channels in $\{P_s\}_{s=1}^S$ is
achieved with equiprobable inputs. In cases where the parallel
channels are not symmetric, a similar result can be shown where the
capacities are replaced with the
mutual information obtained with equiprobable inputs.
\end{remark}

\subsection{MDS codes} \label{subsection:MDSCodes}

In this section some basic properties of MDS codes are provided. For
complete details and proofs, the reader is referred
to~\cite{MacSloanBook} and~\cite{RonBook}.

\begin{definition} \label{definition:MDScodes}
An $(n,k)$ linear block code $\mathcal{C}$ whose minimum distance is
$d$ is called a maximum distance separable (MDS) code if
\begin{equation} \label{equation:MDSdefinition}
d = n - k + 1.
\end{equation}
\end{definition}

\begin{remark}
The RHS of \eqref{equation:MDSdefinition} is the Singleton bound on
the minimum distance of a linear block code.
\end{remark}

\begin{example}[\textbf{MDS codes}]
The $(n,1)$ repetition code, $(n,n-1)$ single parity-check (SPC) code, and
the whole space of vectors over a finite field are all MDS
codes.
\end{example}

The following properties of MDS codes are of interest in the
continuation of this paper:

\begin{proposition}[\textbf{On the generator matrix of an MDS code}]
{\em Let $\mathcal{C}$ be an MDS code of dimension $k$. Then, every
$k$ columns of the generator matrix of $\mathcal{C}$ are linearly
independent.}
\end{proposition}

\begin{corollary} \label{corollary:MDScodewordDetermination}
{\em Every $k$ symbols of a codeword in an MDS code of dimension $k$ completely
characterize the codeword.}
\end{corollary}

Let $S>0$ be an integer number and fix an integer $m>0$ such that
$2^m-1 \geq S$. In the following, we explain how to construct an
MDS code of block length $S$ and a dimension $k\in[S]$. For every
$k \in [2^m-1]$, there exists a $(2^m-1, k)$ RS code over the
Galois field GF$(2^m)$. Every RS code is an MDS
code~\cite[Proposition~4.2]{RonBook}. Two alternatives are
suggested:
\begin{enumerate}
\item \textbf{Shortened RS codes:} Consider an $(2^m-1, k)$ RS code over the Galois field
GF$(2^m)$. Deleting $2^m-1-S$ columns from the generator matrix of
the considered code results in an $(S,k)$ linear block code over the
same alphabet.
The resulting code is an $(S, k)$ MDS code over GF$(2^m)$.
\item \textbf{Generalized RS (GRS) codes:} GRS codes are MDS
codes
which can be constructed over GF$(2^m)$
for every block length $S$ and dimension $k$ (as long as $2^m-1 \geq
S$).
\end{enumerate}

\begin{remark}[\textbf{On the determination of codewords in RS and
GRS codes}] Our main interest in MDS codes is due to
Corollay~\ref{corollary:MDScodewordDetermination}. This property
is even more appealing for the case of RS or GRS codes because the
determination of a codeword in RS or GRS codes is based on a
polynomial interpolation over finite fields (see, e.g.,
\cite[p.~151]{RonBook}).
\end{remark}

\section{The Proposed Coding Scheme (degraded channels)}
\label{section:TheProposedScheme}

In this section, two parallel polar coding schemes are provided
for a set of binary-input, memoryless, degraded and symmetric
parallel channels. First, the simple particular case of $S=3$
parallel degraded channels is studied in
Section~\ref{section:ThreeChannels}. Next, the intermediate case
of arbitrary number of degraded channels is introduced and studied
in Section~\ref{section:parallelEncoding}.

\subsection{Parallel polar coding for $S=3$ degraded channels}
\label{section:ThreeChannels}

Assume that a parallel coding scheme is applied for communication over a set of 3
parallel channels $P_1$, $P_2$, and $P_3$, whose capacities are $C_1
> C_2 > C_3$, respectively. According to
Theorem~\ref{theorem:capacityOfParallelChannels}, the capacity
$C_{\Pi}$ in this case satisfies
\begin{equation*}
C_{\Pi} = C_1 + C_2 + C_3.
\end{equation*}
Fix the rates $R_1 > R_2 > R_3$, satisfying $R_s < C_s$ for all
$s\in[3]$, and let
\begin{equation*}
R \triangleq R_1 + R_2 + R_3.
\end{equation*}
In the following, a parallel polar coding scheme of rate $R$ is described that
achieves reliable communications. Therefore, the proposed scheme achieves the
capacity $C_{\Pi}$ by selecting the rates $R_1,$ $R_2$, and $R_3$ to
be close, respectively, to $C_1$, $C_2$, and $C_3$, and satisfy the
above condition for the rate triple.

Let $\{\mathcal{A}_n^{(s)}\}$ be the information set sequences as in
Corollary~\ref{corollary:informationSetsDegradedChannels}. Fix a
block length $n$, let
\begin{equation*}
k_s \triangleq |\mathcal{A}_n^{(s)}|,\ \  s \in [3]
\end{equation*}
and
\begin{equation*}
k \triangleq k_1 + k_2 + k_3.
\end{equation*}
The encoding of $k$ information bits to 3 codewords: $\mathbf{x}_1$,
$\mathbf{x}_2$, and $\mathbf{x}_3$ is defined. First, the
information bits are arbitrarily partitioned to three groups of
sizes $k_1$, $k_2$ and $k_3$. Next, the encoding of the first two
codewords is performed as follows:
\begin{itemize}
\item The $k_1$ information bits used to encode $\mathbf{x}_1$ are
(arbitrarily) partitioned into three subsets: $\mathbf{u}_{1,1} \in
\mathcal{X}^{k_3}$, $\mathbf{u}_{1,2} \in \mathcal{X}^{k_2-k_3}$,
and $\mathbf{u}_{\text{r}} \in \mathcal{X}^{k_1 - k_2}$.
\item The $k_2$ information bits used to encode $\mathbf{x}_2$ are
(arbitrarily) partitioned into two subsets: $\mathbf{u}_{2,1} \in
\mathcal{X}^{k_3}$ and $\mathbf{u}_{2,2} \in \mathcal{X}^{k_2-k_3}$.
In addition, $\mathbf{u}_{\text{r}}$ (used for encoding
$\mathbf{x}_1$) is also involved in the encoding of $\mathbf{x}_2$.
\item The codewords $\mathbf{x}_1$ and $\mathbf{x}_2$ are defined
similarly to the case of $S=2$ parallel channels. Specifically, in
terms of coset codes:
\begin{eqnarray}
\mathbf{x}_1 = \mathbf{u}_{1,1} G_n \left( \mathcal{A}_n^{(3)}
\right) + \mathbf{u}_{1,2} G_n \left( \mathcal{A}_n^{(2)}
\setminus \mathcal{A}_n^{(3)} \right) + \mathbf{u}_{\text{r}} G_n
\left( \mathcal{A}_n^{(1)} \setminus \mathcal{A}_n^{(2)}\right) +
\mathbf{b} G_b\left( [n] \setminus \mathcal{A}_n^{(1)} \right)
\label{equation:FirstCodeword3Channel} \\
\mathbf{x}_2 = \mathbf{u}_{2,1} G_n \left( \mathcal{A}_n^{(3)}
\right) + \mathbf{u}_{2,2} G_n \left( \mathcal{A}_n^{(2)}
\setminus \mathcal{A}_n^{(3)} \right) + \mathbf{u}_{\text{r}} G_n
\left( \mathcal{A}_n^{(1)} \setminus \mathcal{A}_n^{(2)}\right) +
\mathbf{b} G_b \left( [n] \setminus \mathcal{A}_n^{(1)} \right)
\label{equation:SecondCodeword3Channel}
\end{eqnarray}
where $\mathbf{b} \in \mathcal{X}^{n-k}$ is a predetermined and
fixed vector.
\end{itemize}

The encoding of the codeword $\mathbf{x}_3$ is based on the remaining $k_3$
information bits, denoted by $\mathbf{u}_3 \in \mathcal{X}^{k_3}$.
In addition, the information bits in $\mathbf{u}_{1,2}$,
$\mathbf{u}_{2,2}$ and $\mathbf{u}_{\text{r}}$ are also involved in
the encoding of $\mathbf{x}_3$:
\begin{equation*}
\mathbf{x}_3 = \mathbf{u}_3 G_n \left(\mathcal{A}_n^{(3)}\right) +
\left(\mathbf{u}_{1,2} + \mathbf{u}_{2,2}\right) G_n \left(
\mathcal{A}_n^{(2)} \setminus \mathcal{A}_n^{(3)} \right) +
\mathbf{u}_{\text{r}} G_n \left( \mathcal{A}_n^{(1)} \setminus
\mathcal{A}_n^{(2)}\right) + \mathbf{b} G_n \left([n] \setminus
\mathcal{A}_n^{(1)}\right).
\end{equation*}
Note that the repetition approach is also done for the indices in
$[n]\setminus \mathcal{A}_n^{(2)}$. However, a different approach is
applied to the indices in $\mathcal{A}_n^{(2)} \setminus
\mathcal{A}_n^{(3)}$. The bits corresponding to these indices are
set using a symbol-wise parity-check of $\mathbf{u}_{1,2}$ and
$\mathbf{u}_{2,2}$.

The order of decoding the information bits for all possible
assignments of codewords over a set of three parallel channels is
provided in Table~\ref{table:CodewordAssignments3Channels}. The
decoding starts with the channel $P_1$ with the maximal capacity
$C_1$. Irrespectively of the actual codeword that is transmitted
over $P_1$, the bits which correspond to the indices in
$\mathcal{A}_n^{(1)}$ are decoded using the standard polar
successive cancellation decoding. The decoded bits depend on the
actual codeword which is transmitted over $P_1$. Next, the
decoding proceeds to process the vector observed at the output of
the channel $P_2$, whose capacity is $C_2$. The decoding of
$|\mathcal{A}_n^{(2)}|$ information bits is established in this
decoding step. Note that for a standard successive cancellation
decoding procedure, $n-|\mathcal{A}_n^{(2)}|$ predetermined and
fixed bits are required for proper operation. For the case at
hand, these bits are not all predetermined and fixed. The vector
$\mathbf{b}$ is predetermined, but the rest depends on the
repetition bits $\mathbf{u}_{\text{r}}$. Since the bits
$\mathbf{u}_{\text{r}}$ were decoded at the previous decoding
stage (based on the observation vector of $P_1$), they can be
treated as if they are predetermined and fixed for the decoding of
$\mathbf{x}_2$. Consequently, $|\mathcal{A}_n^{(2)}|$ information
bits are decoded (depending on the actual codeword transmitted
over the channel $P_2$). Finally, the decoding proceeds for the
vector received at the output of the channel $P_3$. As in the
previous decoding steps, the polar successive cancellation
decoding is applied where the bits corresponding to the split
channels indexed by $[n]\setminus\mathcal{A}_n^{(3)}$ are not all
predetermined and fixed (as in contrast to the standard single
channel case). Nevertheless, these bits can be all determined
using the information bits decoded in the two first steps. The
bits in $\mathbf{b}$ are predetermined and fixed. The repetition
bits in $\mathbf{u}_{\text{r}}$ are already available after the
decoding of the information transmitted over $P_1$. The rest, can
be evaluated by taking a bit-wise exclusive-or (xor) of the bits
decoded in the two previous steps. As an example, a combination
shown in Table~\ref{table:CodewordAssignments3Channels} is
described explicitly. Consider the case where the codeword
$\mathbf{x}_2$ is transmitted over the channel $P_1$, and the
codeword $\mathbf{x}_3$ is transmitted over the channel $P_2$. At
the first decoding step, the vectors $\mathbf{u}_{2,1}$,
$\mathbf{u}_{2,2}$ and $\mathbf{u}_{\text{r}}$ are decoded (where
the predetermined bits refer to the vector $\mathbf{b}$). Next,
the vectors $\mathbf{u}_{3}$, and
$\mathbf{u}_{1,2}+\mathbf{u}_{2,2}$, are decoded (the pretermitted
bits for this decoding stage refer to $\mathbf{b}$ and
$\mathbf{u}_{\text{r}}$). After this stage, the information bits
$\mathbf{u}_{1,2}$ can be determined by $ \mathbf{u}_{2,2} +
\left(\mathbf{u}_{1,2}+\mathbf{u}_{2,2}\right)$. Moreover, the
information bits $\mathbf{u}_{1,2}$ are used for the last decoding
stage as predetermined and fixed bits (together with the vectors
$\mathbf{u}_{\text{r}}$ and $\mathbf{b}$). After the last decoding
stage the vector $\mathbf{u}_{1,1}$ is decoded, and the decoding
of all the information bits is completed.

\begin{table*}
\center {\center
\begin{tabular}{|c|c|c|c|c|c|}
\hline \multicolumn{2}{|c|}{\textbf{Channel} $P_1$} &
\multicolumn{2}{|c|}{\textbf{Channel} $P_2$} &
\multicolumn{2}{|c|}{\textbf{Channel} $P_3$} \\
\hline \textbf{Transmitted} & \textbf{Decoded} &
\textbf{Transmitted} & \textbf{Decoded } & \textbf{Transmitted} &
\textbf{Decoded}\\
\textbf{Codeword}  & \textbf{Information}  &
\textbf{Codeword} & \textbf{Information} & \textbf{Codeword}  & \textbf{Information} \\
\hline $\mathbf{x}_1$ & $\mathbf{u}_{1,1}$, $\mathbf{u}_{1,2}$,
$\mathbf{u}_{\text{r}}$ & $\mathbf{x}_2$ & $\mathbf{u}_{2,1}$, $\mathbf{u}_{2,2}$ & $\mathbf{x}_3$ & $\mathbf{u}_3$ \\
\cline{3-6}  & & $\mathbf{x}_3$ & $\mathbf{u}_3$, $\mathbf{u}_{1,2}
+ \mathbf{u}_{2,2}$ & $\mathbf{x}_2$ & $\mathbf{u}_{2,1}$ \\
\hline $\mathbf{x}_2$ & $\mathbf{u}_{2,1}$, $\mathbf{u}_{2,2}$,
$\mathbf{u}_{\text{r}}$ & $\mathbf{x}_1$ & $\mathbf{u}_{1,1}$,
$\mathbf{u}_{1,2}$ & $\mathbf{x}_3$ & $\mathbf{u}_3$ \\ \cline{3-6}
& & $\mathbf{x}_3$ & $\mathbf{u}_3$, $\mathbf{u}_{1,2} +
\mathbf{u}_{2,2}$ & $\mathbf{x}_1$ & $\mathbf{u}_{1,1}$ \\ \hline
$\mathbf{x}_3$ & $\mathbf{u}_{3}$, $\mathbf{u}_{1,2} +
\mathbf{u}_{2,2}$, $\mathbf{u}_{\text{r}}$ & $\mathbf{x}_1$ &
$\mathbf{u}_{1,1}$, $\mathbf{u}_{1,2}$ & $\mathbf{x}_2$ &
$\mathbf{u}_{2,1}$ \\ \cline{3-6} & & $\mathbf{x}_2$ &
$\mathbf{u}_{2,1}$, $\mathbf{u}_{2,2}$ & $\mathbf{x}_1$ &
$\mathbf{u}_{1,1}$ \\
\hline
\end{tabular}}
\caption{The order of decoding the information bits for all possible
assignment of codewords over a set of three parallel channels}
\label{table:CodewordAssignments3Channels}
\end{table*}


\subsection{Parallel polar coding for $S > 3$ degraded channels} \label{section:parallelEncoding}

\subsection*{B.1. Encoding}
A parallel polar encoding is described for the general case. The
technique used for rate-matching encoding
in~\cite{WillemsGorolhovParallelChannels} is incorporated in the
current case as well. This technique is based on MDS codes, in
particular (punctured) RS codes are used in
\cite{WillemsGorolhovParallelChannels} for rate splitting. As
commented in Section~\ref{subsection:MDSCodes}, GRS codes can also
fit for the provided construction. A set of $S-1$ MDS codes over the
Galois field GF$(2^m)$, all with a common block length $S$ are
chosen (either by puncturing an appropriate RS code or using GRS
codes). These codes are denoted by $\mathcal{C}_{\text{MDS}}^{(k)}$,
$k\in[S-1]$, where the code $\mathcal{C}_{\text{MDS}}^{(k)}$ has dimension $k$.

Let $\{P_s\}_{s=1}^S$ be a given set of memoryless degraded and
symmetric parallel channels, whose capacities are ordered such that
$C_1
> C_2
> \cdots
> C_S$. Let
$\{\mathcal{A}_n^{(s)}\}_{s=1}^S$ be the information index sets
satisfying the properties in
Corollary~\ref{corollary:informationSetsDegradedChannels}, for a
block length $n$ and rates $R_1 > R_2 > \cdots > R_{S}$, $R_s <
C_s$, $s\in[S]$. Define
\begin{equation*}
k_s \triangleq |\mathcal{A}_n^{(s)}|, \ \ s \in [S]
\end{equation*}
and
\begin{equation*}
k_{S+1} \triangleq 0.
\end{equation*}
In addition, it is assumed for the purpose of simplicity that $n$ and $k_s$ for all $s \in [S]$,
are integral multiples of $m$. In the provided coding scheme, $k =
\sum_{s=1}^S k_s$ information bits are encoded into $S$ codewords
$\mathbf{x}_s$, $s\in [S]$. As the rates $R_s$, $s\in[S]$ can be
chosen arbitrarily close to $C_s$, respectively, the capacity
$C_{\Pi}$ in~\eqref{equation:capacityParallelChannels} is shown to
be asymptotically achievable (the error performance is considered in
Section~\ref{section:CapacityApproachingProperty}).

\begin{figure}
\centering
\includegraphics[width=3.2in]{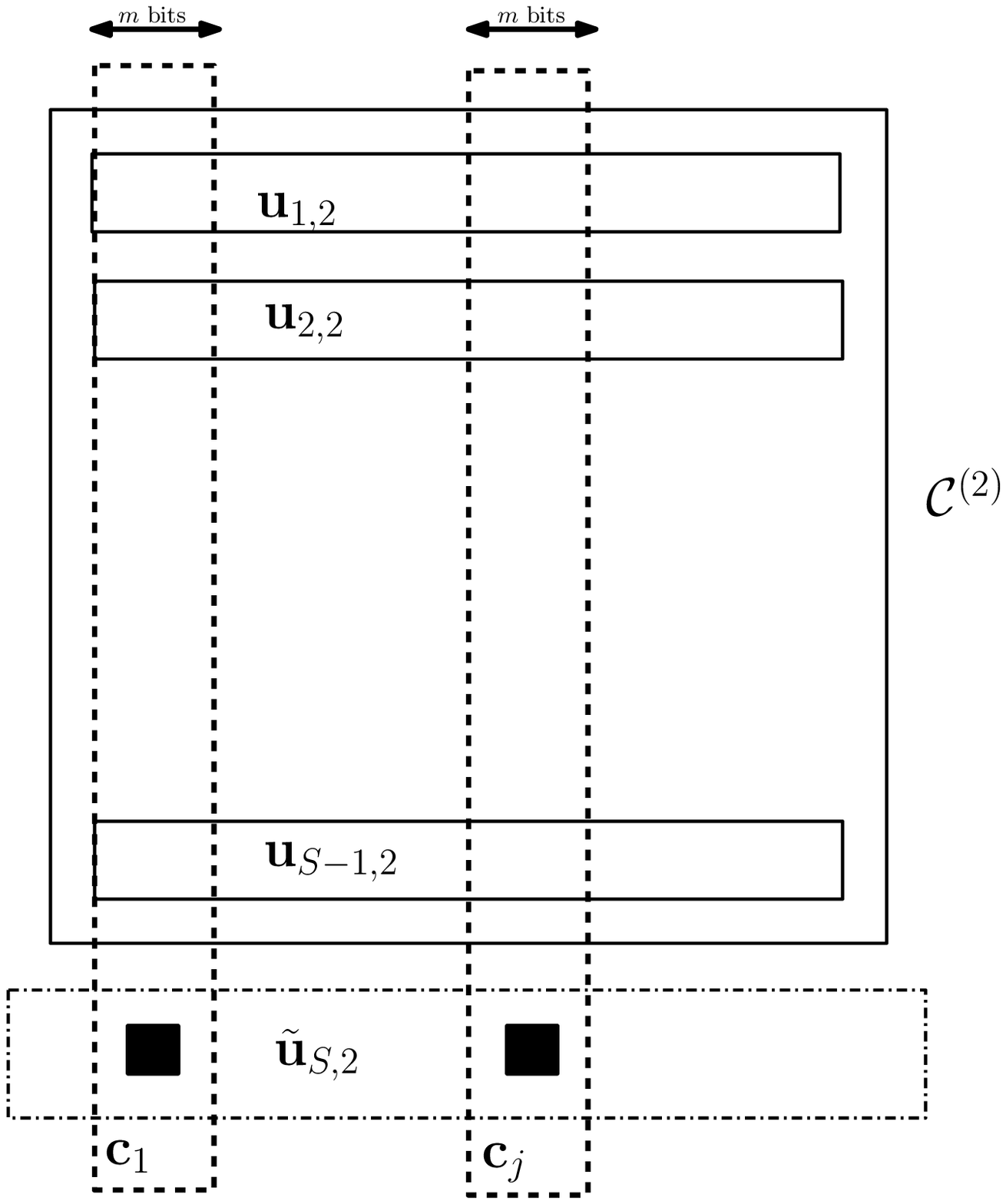}
\caption{\footnotesize{Illustration of the construction of the
vector $\tilde{\mathbf{u}}_{S,2}$. The vectors $\mathbf{u}_{k,s}$,
$k\in[S-1]$ defining the matrix $C^{(2)}$ are shown, along the
columns defining the codewords $\mathbf{c}_j$, $j\in[K_{S-1,S}]$ in
$\mathcal{C}_{\text{MDS}}^{(S-1)}$ }}
 \label{figure:ConstructionStep2}
\end{figure}

Prior to the stage of polar encoding, the $k$ information bits are
first mapped into a set of binary vectors
\begin{equation*}
\mathcal{U} = \left\{\mathbf{u}_{s,l} \in \mathcal{X}^{k_{S-l+1} -
k_{S-l+2}}:\ \ s,l\in[S]\right\}.
\end{equation*}
The $S \cdot k_S$ bits in the vectors $\mathbf{u}_{s,1}$, $s\in[S]$
are plain information bits, chosen arbitrarily from the set of $k$
information bits. The vector set
\begin{align*}
\mathcal{C}_2 \triangleq & \left\{\mathbf{u}_{s,2} =
\bigl(u_{s,2}(1),u_{s,2}(2),\ldots,u_{s,2}(k_{S-1}-k_S)\bigr): \,
s\in[S-1]\right\}
\end{align*}
are also filled with plain information bits, chosen arbitrarily from
the set of remaining $k - S \cdot k_S$ information bits (note that
under the above assumptions $k - S \cdot k_S > 0$). Next, the vector
$\mathbf{u}_{S,2}$ is determined (the following steps are
accompanied with the illustration in
Figure~\ref{figure:ConstructionStep2}):
\begin{enumerate}
\item Each vector in $\mathcal{C}_2$ is rewritten as a row vector of a matrix over
GF$(2^m)$ (this step is illustrated in
Figure~\ref{figure:ConstructionStep2} where each vector is
represented with a horizontal rectangle). Each $m$ consecutive bits
are mapped into a symbol over GF$(2^m)$. This results in the $(S-1)
\times K_{S-1,S}$ matrix over GF$(2^m)$
\begin{equation*}
C^{(2)} = \left(C^{(2)}_{i,j}\right),\ \ \ i\in[S-1],\ j\in
[K_{S-1,S}]
\end{equation*}
where
\begin{equation*}
K_{S-1,S} \triangleq \frac{k_{S-1} - k_S}{m}.
\end{equation*}
The element $C^{(2)}_{i,j}$ is the symbol over GF$(2^m)$
corresponding to the binary length-$m$ vector
\begin{align*}
\Bigl(\mathbf{u}_{i,2}\bigl((j-1)m+1\bigr),\mathbf{u}_{i,2}\bigl((j-1)m+2\bigr),\ldots,
, \mathbf{u}_{i,2}\bigl(jm\bigr)\Bigr)
\end{align*}
where $i\in[S-1]$ and $j\in[K_{S-1,S}]$.
\item Each one of the columns of $C^{(2)}$ are considered as the first
$S-1$ symbols of a codeword in the code
$\mathcal{C}_{\text{MDS}}^{(S-1)}$. These columns are illustrated
with dashed vertical rectangles in
Figure~\ref{figure:ConstructionStep2}. Consequently, these columns
completely determine the codewords
\begin{equation*}
\{\mathbf{c}_j:\ \
j\in[K_{S-1,S}]\}
\end{equation*}
in the MDS $[S,S-1]$ code
$\mathcal{C}_{\text{MDS}}^{(S-1)}$.
\item A length-$K_{S-1,S}$ vector $\tilde{\mathbf{u}}_{S,2}$ over GF$(2^m)$ is defined using the last
symbol of each of the codewords $\mathbf{c}_j$, $j\in[K_{S-1,S}]$,
evaluated in the last step. Each of these symbols is illustrated as
a filled black square in Figure~\ref{figure:ConstructionStep2}.
\item The vector $\mathbf{u}_{S,2}$ is defined by the binary representation of
the vector $\tilde{\mathbf{u}}_{S,2}$ where each symbol over
GF$(2^m)$ is replaced by its corresponding binary length-$m$ vector.
\end{enumerate}

The definition of the remaining vectors in $\mathcal{U}$ continues
in a similar way. Let $2 < l \leq S$, and assume that the vectors
$\mathbf{u}_{s,l'}$ are already defined for all $s\in[S]$ and
$l'<l$, based on
\begin{equation*}
\sum_{s=1}^{l'} (S-(s-1))(k_{S-(s-1)} - k_{S-(s-2)})
\end{equation*}
information bits (from a total of $k$ information bits). The
construction phase for the vectors $\mathbf{u}_{s,l}$, $s\in[S]$ is
defined as follows:
\begin{enumerate}
\item The binary vector set
\begin{equation*}
\mathcal{C}_l = \left\{\mathbf{u}_{s,l}:\ 1\leq s \leq
S-(l-1)\right\}
\end{equation*}
are filled with
\begin{equation*}
\bigl(S-\left(l-1\right)\bigr)\left(k_{S-(l-1)}-k_{S-(l-2)}\right)
\end{equation*}
arbitrarily chosen information bits, out of the remaining
\begin{equation*}
k-\sum_{s=1}^{l'} \bigl(S-(s-1)\bigr)\left(k_{S-(s-1)} -
k_{S-(s-2)}\right)
\end{equation*}
information bits.
\item Each vector in $\mathcal{C}_l$ is rewritten over GF$(2^m)$ as
a row vector in an $(S-(l-1)) \times K_{S-(l-1),S-(l-2)}$ matrix
over GF$(2^m)$
\begin{equation*}
C^{(l)} = \left(C^{(l)}_{i,j}\right)
\end{equation*}
where
\begin{equation*}
K_{S-(l-1),S-(l-2)} \triangleq \frac{k_{S-(l-1)} - k_{S-(l-2)}}{m}
\end{equation*}
and $C^{(l)}_{i,j}$, $i\in[S-(l-1)]$, $j\in[K_{S-(l-1),S-(l-2)}]$,
equals the symbol in GF$(2^m)$ corresponding to the binary
length-$m$ vector
\begin{align*}
&
\Bigl(\mathbf{u}_{i,l}\bigl((j-1)m+1\bigr),\mathbf{u}_{i,l}\bigl((j-1)m+2\bigr),
\ldots,\mathbf{u}_{i,2}\bigl(jm\bigr)\Bigr).
\end{align*}
\item Each column in $C_l$ is a vector of $S-(l-1)$ symbols over
GF$(2^m)$. Hence, it completely determines a codeword $\mathbf{c}_j
= (c_{j,1}, c_{j,2},\ldots, c_{j,S})$, $j\in[K_{S-(l-1),S-(l-2)}]$,
in the MDS $[S,S-(l-1)]$ code
$\mathcal{C}_{\text{MDS}}^{(S-(l-1))}$. The columns of $C_l$ are
considered as the first $S-(l-1)$ symbols of a codeword in the code
$\mathcal{C}_{\text{MDS}}^{(S-(l-1))}$.
\item Evaluate the remaining symbols for each of the codewords
$\mathbf{c}_j$, $j\in[K_{S-(l-1),S-(l-2)}]$.
\item The length-$K_{S-(l-1),S-(l-2)}$ vectors $\tilde{\mathbf{u}}_{s,l} =
(\tilde{u}_{s,l}(1),\ldots,\tilde{u}_{s,l}(K_{S-(l-1),S-(l-2)}))$,
$s > S-(l-1)$, over GF$(2^m)$ are defined using the codewords
$\mathbf{c}_j$, $j\in[K_{S-(l-1),S-(l-2)}]$ according to
\begin{equation*}
\tilde{u}_{s,l}(j) = c_{j,s}.
\end{equation*}
\item
For every $s>S-(l-1)$, The vector $\mathbf{u}_{s,l}$ is defined to
be the binary representation of the vector
$\tilde{\mathbf{u}}_{s,l}$ (where each symbol over GF$(2^m)$ is
replaced with its binary length-$m$ vector representation).
\end{enumerate}

The parallel polar codewords are defined using the coset code
notation. Specifically, the codewords $\mathbf{x}_s$, $s\in[S]$, are
defined according to
\begin{equation}
\mathbf{x}_s = \sum_{l=1}^S \mathbf{u}_{s,l} G_n
\left(\mathcal{A}_n^{(S-(l-1))} \setminus
\mathcal{A}_n^{(S-(l-2))}\right) + \mathbf{b} G_n \left([n]
\setminus \mathcal{A}_n^{(1)}\right),\ \ \ s\in[S]
\label{equation:parallelPolarEncoding}
\end{equation}
where $\mathcal{A}_n^{(S+1)} \triangleq \emptyset$ and $\mathbf{b}
\in \mathcal{X}^{n-k_1}$ is a binary predetermined and fixed vector.

\subsection*{B.2. Decoding} \label{section:parallelDecoding}

The decoding process starts with the observations received at the
output of the channel $P_1$ whose capacity is maximal. Assume that
the codeword $x_{\pi^{-1}(1)}$ is transmitted over $P_1$. A polar
successive cancellation decoding, with respect to the information
index set $\mathcal{A}_n^{(1)}$, is applied to the received vector.
This allows the decoding of the vectors $\mathbf{u}_{\pi^{-1}(1),
l}$, $l\in[S]$ (as if they are the information bits of the
considered polar code). If $\pi^{-1}(1) = 1$, then indeed all the
vectors $\mathbf{u}_{\pi^{-1}(1), l} = \mathbf{u}_{1,l}$, $l\in[S]$
are information bit vectors. Generally, only a subset of these
vectors comprise of information bits, the rest are coded binary
representation of coded symbols of the chosen MDS codes.

At the second stage, the decoding of the received vector over $P_2$,
which denotes probability transition of the channel with the second
largest capacity, is concerned. Assume that the codeword
$\mathbf{x}_{\pi^{-1}(2)}$ is transmitted over $P_2$. A polar
successive cancellation decoding is used. This decoding procedure is
capable of decoding $|\mathcal{A}_n^{(2)}|$ bits based on
$n-|\mathcal{A}_n^{(2)}|$ predetermined and fixed bits. For the
current decoding procedure, $n-|\mathcal{A}_n^{(1)}|$ of these bits
are the predetermined and fixed bits in $\mathbf{b}$. The rest of
$|\mathcal{A}_n^{(1)}|-|\mathcal{A}_n^{(2)}|$ bits are based on the
bits decoded at the previous decoding stage. Specifically, the bit
vector $\mathbf{u}_{\pi^{-1}(2),S}$ can be evaluated using the bit
vector $\mathbf{u}_{\pi^{-1}(1),S}$. Recall that
$\mathbf{u}_{\pi^{-1}(2),S}$ is the binary representation of
$\tilde{\mathbf{u}}_{\pi^{-1}(2),S}$. Moreover, each of the symbols
of $\tilde{\mathbf{u}}_{\pi^{-1}(2),S}$ belongs to a codeword in the
$[S,1]$ MDS code $\mathcal{C}_{\text{MDS}}^{(1)}$. These codewords
are fully determined from the vector $\mathbf{u}_{\pi^{-1}(1),S}$ as
follows:
\begin{enumerate}
\item Rewrite the vector $\mathbf{u}_{\pi^{-1}(1),S}$ over GF$(2^m)$
where each consecutive $m$ bits are rewritten by the corresponding
symbol over GF$(2^m)$. Denote by
\begin{equation*}
\tilde{\mathbf{u}}_{\pi^{-1}(1),S} = \bigl(
\tilde{u}_{\pi^{-1}(1),S}(1),\ldots,\tilde{u}_{\pi^{-1}(1),S}(K_{1,2})\bigr)
\end{equation*}
the resulting length-$K_{1,2}$ vector over GF$(2^m)$.
\item For each symbol $\tilde{u}_{\pi^{-1}(1),S}(j)$, $j\in [K_{1,2}]$, find
the codeword
\begin{equation*}
\mathbf{c}_j = (c_{j,1},\ldots,c_{j,S}) \in
\mathcal{C}_{\text{MDS}}^{(1)}
\end{equation*}
whose $\pi^{-1}(1)$-th symbol satisfies $c_{j,\pi^{-1}(1)} =
\tilde{u}_{\pi^{-1}(1),S}(j)$. These codewords are fully determined
by the considered symbols.
\item Define the vector
\begin{equation*}
\tilde{\mathbf{u}}_{\pi^{-1}(2),S} =
\left(\tilde{u}_{\pi^{-1}(2),S}(1),\ldots,\tilde{u}_{\pi^{-1}(2),S}(K_{1,2})\right)
\end{equation*}
according to $\tilde{u}_{\pi^{-1}(2),S}(j) = c_{j,\pi^{-1}(2)}$ for every
$j\in \left[K_{1,2}\right]$.
\item The vector
\begin{equation*}
\mathbf{u}_{\pi^{-1}(2),S} =
\left(u_{\pi^{-1}(2),S}(1),\ldots,u_{\pi^{-1}(2),S}(k_1 -
k_2)\right)
\end{equation*}
is set to the binary representation of
$\tilde{\mathbf{u}}_{\pi^{-1}(2),S}$. That is, the bits
$u_{\pi^{-1}(2),S}((j-1)m+1),\ldots,u_{\pi^{-1}(2),S}(jm)$ are the
binary representation of the symbol $\tilde{u}_{\pi^{-1}(2),S}(j)
\in \text{GF}(2^m)$, $j\in K_{1,2}$.
\end{enumerate}

With both $\mathbf{b}$ and $\mathbf{u}_{\pi^{-1}(2),S}$ as
predetermined and fixed bits, the polar successive cancellation
decoding can be applied. Consequently, after the second decoding
stage, all the $S$ binary vectors $\mathbf{u}_{\pi^{-1}(2),s}$,
$s\in[S]$, are fully determined. Moreover, based on the codewords
$\mathbf{c}_j$, $j\in [K_{1,2}]$, the vectors
$\mathbf{u}_{\pi^{-1}(s),S}$, are fully determined for all $s\geq 2$
as well.

Next, the remaining $S-2$ decoding stages are described. It is
assumed that after the $(s-1)$-th decoding stage, where $2<s<S$, the vectors
$\mathbf{u}_{\pi^{-1}(s'),l}$ for either $1\leq s' < s$ and $l \in
[S]$, or $s'\geq s$ and $S-s+3\leq l \leq S$, were decoded at
previous stages. At the $s$-th stage, the decoding is extended for
the vectors $\mathbf{u}_{\pi^{-1}(s),l}$ for all $l\in[S]$ and the
vectors $\mathbf{u}_{\pi^{-1}(s'),S-s+2}$ for all $s'\in[S]$.

In order to apply the polar successive cancellation decoding
procedure to the vector received over the channel $P_s$, the bits in
$\mathbf{b}$ and $\{\mathbf{u}_{\pi^{-1}(s),l}\}_{l\geq S-(s-2)}$
must be known for the procedure. The vector $\mathbf{b}$ is clearly
known. In addition, the bits in
$\{\mathbf{u}_{\pi^{-1}(s),l}\}_{l\geq S-(s-3)}$ are already decoded
in previous stages. It is left to determine the bits in
$\mathbf{u}_{\pi^{-1}(s), S-(s-2)}$. These bits are determined in a
similar manner as in the decoding stage for $s=2$, where the vector
$\mathbf{u}_{\pi^{-1}(2),S}$ is determined. Moreover, the
determination of $\mathbf{u}_{\pi^{-1}(s), S-(s-2)}$ is established
along with the determination of $\mathbf{u}_{\pi^{-1}(s'), S-(s-2)}$
for all $s'\geq s$, in the following way:
\begin{enumerate}
\item The binary vectors $\mathbf{u}_{\pi^{-1}(s'),S-s+2}$ for $s'<
s$ are already decoded at previous stages. Rewrite these vectors
over GF$(2^m)$ where each consecutive $m$ bits are rewritten by the
corresponding symbol over GF$(2^m)$. Denote the set of resulting
vectors by
\begin{equation*}
\mathcal{D} = \left\{\tilde{\mathbf{u}}_{\pi^{-1}(s'),S-s+2} =
\bigl(\tilde{u}_{\pi^{-1}(s'),S-s+2}(1),\ldots,\tilde{u}_{\pi^{-1}(s'),S-s+2}(K_{s-1,s})\bigr):
\, s' < s\right\}.
\end{equation*}
\item The set $\mathcal{D}$ completely describes $K_{s-1,s}$
codeword $\mathbf{c}_j = (c_{j,1},\ldots, c_{j,S})$, $j\in
[K_{s-1,s}]$, all in the code $\mathcal{C}_{\text{MDS}}^{(s-1)}$ and
satisfy the constraints:
\begin{equation}
c_{j,\pi^{-1}(s')} = \tilde{u}_{\pi^{-1}(s'),S-s+2}(j),\ \ 1\leq s'
< s.
\end{equation}
\item Define the vectors
\begin{equation*}
\tilde{\mathbf{u}}_{\pi^{-1}(s'),S-s+2} = \bigl(
\tilde{u}_{\pi^{-1}(s'),S-s+2}(1), \ldots,
\tilde{u}_{\pi^{-1}(s'),S-s+2}(K_{s-1,s})\bigr)
\end{equation*}
for all $s' \geq s$ by
\begin{equation*}
\tilde{u}_{\pi^{-1}(s'),S-s+2}(j) \triangleq c_{j,\pi^{-1}(s')}, \ \
j\in[K_{s-1,s}].
\end{equation*}
\item The vectors $\mathbf{u}_{\pi^{-1}(s'),S-s+2}$ are determined
for all $s'\geq s$ by the binary representation of
$\tilde{\mathbf{u}}_{\pi^{-1}(s'),S-s+2}$.
\end{enumerate}

Based on successive cancellation at the current decoding stage, the
$k_s$ bits corresponding to the information set
$\mathcal{A}_n^{(s)}$ are decoded. This completes the decoding of
all the binary vectors $\mathbf{u}_{\pi^{-1}(s),l}$ for $l\in[S]$.

\begin{remark}[\textbf{On channels with equal
capacities}] The case where for an index $s'\in[S]$, $C_{s'} =
C_{s'+1}$ is treated by skipping the construction of
$\mathcal{C}_{s'}$. The coset codewords are defined by
\begin{align*} \label{equation:parallelPolarEncodingEqualCapacitiesCase}
\mathbf{x}_s & = \sum_{l=1}^{s'-1} \mathbf{u}_{s,l} G_n
\left(\mathcal{A}_n^{S-(l-1)} \setminus
\mathcal{A}_n^{S-(l-2)}\right) \\
& \ \ \ + \mathbf{u}_{s,s'+1} G_n \left(\mathcal{A}_n^{(S-s')}
\setminus
\mathcal{A}_n^{(S-s'+2)}\right) \\
& \ \ \ + \sum_{l=s'+2}^{S} \mathbf{u}_{s,l} G_n
\left(\mathcal{A}_n^{(S-(l-1))} \setminus
\mathcal{A}_n^{(S-(l-2))}\right) \nonumber\\
& \ \ \ + \mathbf{b} G_n \left([n]
\setminus \mathcal{A}_n^{(1)}\right),\ \ \ s\in[S]
\end{align*}
At the decoding stage, two consecutive polar successive cancellation
decoding can be performed for both vectors received at the output of
the channel $P_{s'}$ and $P_{s'+1}$.
\end{remark}

\subsection*{B.3. Capacity-approaching property} \label{section:CapacityApproachingProperty}

\begin{theorem} \label{theorem:capacityApproachingProperty}
{\em The provided parallel coding scheme achieves the capacity of
every arbitrarily-permuted memoryless degraded and symmetric set
of parallel channels.}
\end{theorem}

\begin{proof}
Consider a set of $S$ arbitrary-permuted degraded memoryless
parallel channels $P_s$, $s\in[S]$, whose capacities are $C_s$,
$s\in[S]$, respectively, and assume that the channels are ordered so that
\begin{equation*}
C_1 \geq C_2 \geq \cdots \geq C_{S}.
\end{equation*}
According to Theorem~\ref{theorem:capacityOfParallelChannels}, the
capacity $C_{\Pi}$ for the considered model is equal to the sum
in~\eqref{equation:capacityParallelChannels}. For a rate
$R<C_{\Pi}$, choose a rate set $\{R_s\}_{s=1}^S$ satisfying
\begin{equation}
R_s < C_s, \quad \sum_{s=1}^S R_s \geq R.
\label{equation:sumRates}
\end{equation}
The parallel polar coding in Section~\ref{section:parallelEncoding}
is considered. The rate of the proposed scheme is given by
\begin{equation*}
\frac{1}{n}\sum_{s=1}^S |\mathcal{A}_n^{(s)}|.
\end{equation*}
From~\eqref{equation:rateProperties}
and~\eqref{equation:sumRates}, it follows that the proposed scheme
can be designed to operate at every rate below capacity. It is
left to prove that the block error probability of the proposed
scheme can be made arbitrarily small for a sufficiently large
block length.

Consider the vectors
\begin{equation} \label{equation:likeInformationBitVectors}
\mathbf{u}_{s,l},\ \  s,l \in [S]
\end{equation}
in~\eqref{equation:parallelPolarEncoding}. These vectors include all
the information bits to be transmitted (in addition to coded
versions of these bits). These vectors are determined either via the
successive cancellation decoding procedure of the polar codes, or
determined by the MDS code structure applied in the parallel scheme.
The successive cancellation decoding procedure is based on detecting
the input to the set of split channels $P_{s,n}^{(l)}$ where
$s\in[S]$ and $l \in \mathcal{A}_n^{(s)}$. The information bit
corresponding to a split channel $P_{s,n}^{(l)}$, is denoted by
$a_{s,l}$. Note that the bit $a_{s,l}$ is either determined by the
successive cancellation decoding procedure for polar codes, or else
determined by the codeword of an MDS code for which it belongs to.
In cases where the bit $a_{s,l}$ is decoded via a polar successive
cancellation decoding procedure, the decoded bit is denoted by
$\hat{a}_{s,l}$.

The bits decoded via polar successive cancellation decoding
procedure, based on the received vector at the output of the channel
$P_s$, $s\in[S]$, are
\begin{equation} \label{equation:decodedBits}
\hat{a}_{s,l}, \quad l\in\mathcal{A}_n^{(s)}.
\end{equation}
Note that the bits in~\eqref{equation:decodedBits} do not include
all the bits in~\eqref{equation:likeInformationBitVectors}.
Nevertheless, the rest of the bits
in~\eqref{equation:likeInformationBitVectors} are fully determined
from the decoded bits in~\eqref{equation:decodedBits} based on the
MDS code structure (as detailed in the previous section).

Assuming that a permutation $\pi$ is applied to the transmission of
codewords, define the events
\begin{align*}
\mathcal{F}_{s,l} \triangleq & \left\{\hat{a}_{s,l} \neq
a_{\pi^{-1}(s),l},\ \hat{a}_{s',l'} = a_{\pi^{-1}(s'),l'}:\right.\\
&\left. \hspace*{3cm} \text{
for all } s'\leq s, l' < l\right\}
\end{align*}
where $s\in[S]$ and $l\in\mathcal{A}_n^{(s)}$. Since all the
information bits can be fully determined from the bits
in~~\eqref{equation:decodedBits}, the conditional block error
probability is given by
\begin{equation*}
P_{\text{e}|m} = \Pr \left( \cup_{s=1}^S
\cup_{l\in\mathcal{A}_n^{(s)}} \mathcal{F}_{s,l}\right)
\end{equation*}
where $m$ is the transmitted message (representing the $k$
information bits). The events
$\mathcal{E}_l(P_s)$ for $s\in[S]$ and $l\in\mathcal{A}_n^{(s)}$,
defined in~\eqref{equation:eventDefinitionErrorInL}, can be shown to be independent
of the transmitted message \cite{ArikanPolarCodes}. Moreover, it follows that
\begin{equation*}
\mathcal{F}_{s,l} \subseteq \mathcal{E}_l(P_s).
\end{equation*}
Consequently, the average block error probability is upper bounded
using the union bound according to
\begin{equation} \label{equation:unionBound}
P_{\text{e}} \leq \sum_{s\in[S]} \sum_{l \in \mathcal{A}_n^{(s)}}
\Pr\left(\mathcal{E}_l(P_s)\right)
\end{equation}
Finally, plugging the upper bound on the error
probability~\eqref{equation:SplitChannelPerformanceProperty}
into~\eqref{equation:unionBound}, assures that for every fixed
$S>0$, the block error probability can be made arbitrarily low as
the block length increases.
\end{proof}

\begin{remark}[\bf{On symmetry condition for the applied coding scheme}] \label{remark:WhyWeNeedSymmetry}
In order to use the result
in~\eqref{equation:SplitChannelPerformanceProperty} for the proof
of Theorem~\ref{theorem:capacityApproachingProperty}, we rely on
the symmetry result in~\cite{ArikanPolarCodes}. Specifically, it
is shown in~\cite{ArikanPolarCodes} that for symmetric channels
according to Definition~\ref{definition:symmetricBDMC}, the error
performance of the polar coding successive cancellation process is
independent on both the information bits and the predetermined and
fixed bits. This result is of particular importance for our scheme
as the predetermined and fixed bits of the channel polarization
method are not predetermined and fixed in our scheme.
\end{remark}

\section{Parallel Polar Coding for Non-Degraded Parallel Channels}
\label{section:NonDegradedCase}

In this section, a parallel polar coding scheme is provided for transmissions over non-degraded parallel channels. With the introduction of non-degraded channels, the property which must be relaxed is the monotonicity of the information sets in~\eqref{equation:monotonicityProperty4ParallelChannels}. Consequently, a proper modification must be introduced. In fact, it is the ordering of the successive cancellation process which is found to be the key ingredient in dealing with the non-degraded case. That is, the decoding is not carried channel-after-channel as in Section~\ref{section:TheProposedScheme}, but for each bit index a different ordering of channels is applied. If the decoding order is kept channel-after-channel, it can be shown that the decoding method presented in Section~\ref{section:TheProposedScheme} can not achieve capacity. In particular, an upper bound on the capacity of the coding method in Section~\ref{section:TheProposedScheme} is first provided. Next, two alternative coding schemes with modified ordering are presented.

\subsection{Upper bound for channel-after-channel ordering}

\subsection*{A.1. Signaling over Parallel Erasure Channels} \label{section:signalingOnErasure}

The following proposition, provided in~\cite{CompoundCapacity},
considers the Bhattacharyya parameters of the split channels:

\begin{proposition}[\textbf{On the worst Bhattacharyya parameter}~\cite{CompoundCapacity}]
\label{proposition:WorstBConstant} {\em Let $p$ be a binary-input
memoryless output-symmetric channel, and consider the split
channel $p_n^{(l)}$ where $l\in[n]$. Then, among all such
binary-input memoryless output-symmetric channels $p$ whose
Bhattacharyya parameter equals $B$, the binary erasure channel has
the maximal Bhattacharyya parameter $B(p_n^{(l)})$, for every
$l\in[n]$.}
\end{proposition}

The proof of Proposition~\ref{proposition:WorstBConstant} is based
on a tree-channel characterization of split channels, in addition
to an argument which is related to extremes of information
combining. Based on Proposition~\ref{proposition:WorstBConstant},
a polar signaling scheme is provided in~\cite{CompoundCapacity}
for reliable communication in a compound setting. A similar
technique is used in the following for the parallel channel
setting.

Consider the parallel transmission model in
Section~\ref{section:communicationModel}. In this section, it is
assumed that the parallel channels are binary-input memoryless and
symmetric, but are not necessarily degraded. We further assume,
without loss of generality, that the set of parallel channels
$\{P_s\}_{s\in[S]}$, are ordered such that
\begin{equation*}
B(P_1) \leq B(P_2) \leq \ldots \leq B(P_S)
\end{equation*}
where $B(P_s)$ is the Bhattacharayya parameter of the channel $P_s$,
$s\in[S]$ (note that the Bhattacharyya parameter varies from 0 to 1
with the extremes of zero and one for a noiseless and completely
noisy channels, respectively). Next, consider the set of parallel
binary erasure channels, $\{\delta_{s}\}_{s\in[S]}$ where the
erasure probability of the channel $\delta_{s}$ equals $B(P_s)$,
$s\in[S]$. These erasure channels form a family of $S$
stochastically degraded channels. Consequently, based on
Theorem~\ref{theorem:capacityApproachingProperty}, the parallel
polar coding scheme in Section~\ref{section:parallelEncoding}
achieves a rate of $S-\sum_{s=1}^S B(P_s)$ over the set of erasure
channels, under the successive cancellation decoding scheme detailed
in Section~\ref{section:parallelDecoding}. The following corollary
addresses the performance of the same coding scheme over the
original set of parallel channels:

\begin{corollary}{\em The polar coding scheme for the parallel erasure channels,
operates reliably over the original parallel channels.}
\end{corollary}
\begin{IEEEproof}
The suggested coding scheme performs reliably over the parallel binary
erasure channels. The decoding process, as described in
Section~\ref{section:parallelDecoding}, includes a sequence of
successive cancellation decoding operations applied to the polar
codes over each one of the parallel channels. As shown in the
proof of Theorem~\ref{theorem:capacityApproachingProperty},
reliable communication is obtained based on reliably decoding each of the successive cancellation operations. It is therefore required to show that the successive cancellation over the original channels $\{P_s\}_{s\in[S]}$ can also be carried reliably, this follows as a consequence of Proposition~\ref{proposition:WorstBConstant}. Denote the sequences of information sets chosen for reliable communication over the erasure channels $\{\delta_{s}\}_{s\in[S]}$ by $\{\mathcal{A}_n^{(s)}\}_{s\in[S]}$. 
Fix an arbitrary channel $P_s$ from the set of parallel channels, and an arbitrary index $l \in A_n^{(s)}$. Consider next the error event $\mathcal{E}_l(P_s)$ in~\eqref{equation:eventDefinitionErrorInL}. According to~\cite{ArikanPolarCodes}, this error event is upper bounded by
\begin{equation} \label{equation:boundingStep1}
\Pr\bigl(\mathcal{E}_l(P_s)\bigr) \leq B\bigl((P_s)_n^{(l)}\bigr)
\end{equation}
where $B\bigl((P_s)_n^{(l)}\bigr)$ denotes the Bhattacharayya
parameter of the split channel $(P_s)_n^{(l)}$. From
Proposition~\ref{proposition:WorstBConstant}, it follows that
\begin{equation} \label{equation:boundingStep2}
B\bigl((P_s)_n^{(l)}\bigr) \leq B\bigl((\delta_s)_n^{(l)}\bigr)
\end{equation}
where $B\bigl((\delta_s)_n^{(l)}\bigr)$ is the Bhattacharayya
constant of the split channel $(\delta_s)_n^{(l)}$. Fix $0 < \beta
< \frac{1}{2}$ as in~\cite{TelatarArikanPolarRate}.
From~\eqref{equation:boundingStep1}
and~\eqref{equation:boundingStep2}, it follows
from~\cite{TelatarArikanPolarRate} that
\begin{equation*}
\Pr\bigl(\mathcal{E}_l(P_s)\bigr) \leq 2^{-n^\beta}.
\end{equation*}
Consequently, the successive cancellation decoding operations can be carried reliably for each one of the original channels, which completes the proof.
\end{IEEEproof}

\subsection*{A.2. A Compound Interpretation of Monotone Index Set Design and Related Results}

The parallel coding scheme provided in Section~\ref{section:TheProposedScheme} is based on a monotonic sequence of index sets $\{\mathcal{A}_n^{(s)}\}_{s\in[S]}$ satisfying the conditions in Corollary~\ref{corollary:informationSetsDegradedChannels}. As explained in Remark~\ref{remark:onIndices4DegradedChannels}, the index sets in  $\mathcal{A}_n^{(s)}$, $s\in[S]$ are `good' for all the channels $P_{s'}$, $s'\geq s$. Here, as in Remark~\ref{remark:onIndices4DegradedChannels}, `good' means that the corresponding Bhattacharayya parameters of the corresponding split channels satisfy the polarization properties studied in~\cite{ArikanPolarCodes},~\cite{TelatarArikanPolarRate}. The index set sequences $\{\mathcal{A}_n^{(s)}\}_{s\in[S]}$  are applied in this paper to parallel transmission. Even though the compound setting and the problem of parallel transmissions are, at first glance different, the actual problem of finding an index set which is `good' for a set of channels is similar to the problem studied in~\cite{CompoundCapacity} in the compound model.

In the compound setting, the transmission takes place over one channel which belongs to a predetermined set of channels. It is assumed in the current discussion that (only) the receiver knows the channel over which the transmission takes place. If a polar code is applied in such a compound setting, then a suitable index set is required. Such an index set must be `good' for all the channels in the set. The maximal rate over which such a polar coding scheme performs reliably is termed as the compound capacity of polar codes. Obviously, the compound capacity relates to the size of possible `good' index sets.

Upper and lower bounds on the compound capacity of polar codes
under successive cancelation decoding are provided
in~\cite{CompoundCapacity}. These bounds are defined using the
notion of tree-channels. Let $p$ be a binary-input memoryless
output-symmetric channel. For a binary vector of length $k$,
$\mathbf{\sigma}=(\sigma_1,\sigma_2,\ldots,\sigma_k)$, the
tree-channel associated to $\mathbf{\sigma}$ is denoted by
$p^{\mathbf{\sigma}}$. The actual definition of the tree-channel
is not required for the following discussion, and is therefore
omitted (the reader is referred to~\cite{CompoundCapacity} and
references therein for more details). It is noted that the
tree-channel is also binary-input memoryless and output-symmetric.
Moreover, it is further noted in~\cite{CompoundCapacity} that the
tree-channel $p^{\mathbf{\sigma}}$, is equivalent to the
split-channel $p_n^{(l)}$ where $\sigma$ is the binary expansion
of $l$.

Let $\{P_s\}_{s\in[S]}$ be a set of $S$ binary-input memoryless
output-symmetric channels. It is shown in~\cite{CompoundCapacity}
that the compound capacity for the considered setting
$C\bigl(\{P_s\}_{s\in[S]}\bigr)$ is lower bounded by
\begin{equation} \label{equation:compoundCapacityLowerBound}
C\bigl(\{P_s\}_{s\in[S]}\bigr) \geq 1-\frac{1}{2^k} \sum_{\mathbf{\sigma}\in\{0,1\}^k} \max_{s\in[S]} B\bigl(P_s^{\mathbf{\sigma}}\bigr)
\end{equation}
where $k \in \mathbb{N}$ and $B\bigl(P_s^{\mathbf{\sigma}}\bigr)$
is the Bhattacharyya parameter of the tree-channel
$P_s^{\mathbf{\sigma}}$. Moreover, this lower bound is a
constructive bound. That is, the construction of an appropriate
index set sequence $\mathcal{A}_n\bigl(\{P_s\}_{s\in[S]}\bigr)$ is
inherent from the lower bound. The polar code corresponding to
this index set has an asymptotically low decoding error
probability under successive cancellation decoding (for every
channel in the set $\{P_s\}_{s\in[S]}$).

\begin{remark}[\textbf{On the derivation of~\eqref{equation:compoundCapacityLowerBound}}]
\label{remark:onTherivationOfCompundBounds}
The actual derivation in~\cite{CompoundCapacity} is provided for
two channels $P$ and $Q$. Nevertheless, the arguments
in~\cite{CompoundCapacity} are suitable for the case of $S>2$
channels. The proof of the bounds in~\cite{CompoundCapacity} is
based on two major arguments. The first argument consider a
sequential transformations of a given channel $P$ to a sequence of
sets of tree-channels. Initially, the channel $P$ is transformed
into a pair of tree-channels $P^0$ and $P^1$. Next, each of these
tree-channels is transformed again to another pair, and the
transformation repeats recursively. It is shown that instead of
transmitting bits corresponding to indices induced by the
polarization of the original channel $P$, at each transformation
level $k$, the problem is equivalent to transmitting a fraction
$\frac{1}{2^k}$ of the bits based on the indices induced by the
polarization of the corresponding tree channels
$\{P^{\mathbf{\sigma}}\}_{\mathbf{\sigma}\in\{0,1\}^k}$. The first
argument is therefore not affected by the number of channels (as
it concerns a property of a single channel). The second argument
is identical to the more simpler polarization scheme detailed in
Section~\ref{section:signalingOnErasure}. This polarization
scheme, based on binary erasure channels, can be applied to every
set of tree-channels $\{P_s^{\mathbf{\sigma}}\}_{s=1}^{S}$,
$\mathbf{\sigma} \in \{0,1\}^k$. Based on this polarization
scheme, a rate of $\frac{1}{2^k} \left(1-\max_{s\in[S]}
B\bigl(P_s^{\mathbf{\sigma}}\bigr)\right)$ is guaranteed for each
$\mathbf{\sigma}\in\{0,1\}^k$.
\end{remark}

\begin{corollary}[\textbf{Improved parallel polar coding scheme}] \label{corollary:ImprovedPolarCodingScheme}
{\em Consider the transmission over a set of parallel binary-input
memoryless and output-symmetric channels $\{P_s\}_{s\in[S]}$. Fix
an order $P_{s_1},P_{s_2},\ldots,P_{s_{S}}$ of channels and
$k\in\mathbb{N}$. Then, reliable transmission is achievable based
on the parallel polar coding scheme in
Section~\ref{section:TheProposedScheme}, whose rate is given by
\begin{equation} \label{equation:lowerBoundParallelNonDegraded}
C(P_{s_S}) + S - 1 - \frac{1}{2^k}  \sum_{s\in[S-1]} \sum_{\mathbf{\sigma}\in\{0,1\}^k} \max_{i\in\{s,\ldots S\}} B\bigl(P_{s_{i}}^{\mathbf{\sigma}}\bigr).
\end{equation}
}
\end{corollary}

\begin{IEEEproof}
Define the channel sets
\begin{equation*}
\mathcal{P}_s \triangleq \left\{ P_{s_i} \right\}_{i=s}^S, \quad s\in[S].
\end{equation*}
For each channel set $\mathcal{P}_s$, $s\in[S]$, the compound
setting is considered. Based on the lower bound
in~\eqref{equation:compoundCapacityLowerBound} and its associated
index set sequence, a set sequence
$\mathcal{A}_n\bigl(\mathcal{P}_s\bigr)$ exists for every
$s\in[S]$, such that
\begin{equation} \label{equation:rateOfIndexSets4CompoundSetting}
\frac{1}{n}\mathcal{A}_n\bigl(\mathcal{P}_s\bigr) \geq 1-\frac{1}{2^k}
\sum_{\mathbf{\sigma}\in\{0,1\}^k} \max_{i\in\{s,\ldots S\}} B\bigl(P_{s_{i}}^{\mathbf{\sigma}}\bigr)
\end{equation}
and reliable decoding is guaranteed for all the channels in the set
$\mathcal{P}_s$ under successive cancellation decoding. As an
immediate consequence of the construction, for every $n$, the index
sets form a monotonic sequence (i.e., if an index is 'good' for a
set of channels, it must be 'good' for a subset of these channels).
Therefore, the monotone set sequences for the polar construction is
provided and the parallel polar scheme in
Section~\ref{section:TheProposedScheme} can be applied. The rate of
the resulting scheme is given by summing over the rates
in~\eqref{equation:rateOfIndexSets4CompoundSetting} which adds to
\begin{equation*}
S - \frac{1}{2^k}  \sum_{s\in[S]} \sum_{\mathbf{\sigma}\in\{0,1\}^k}
\max_{i\in\{s,\ldots S\}} B\bigl(P_{s_{i}}^{\mathbf{\sigma}}\bigr).
\end{equation*}
Since the last channel set $\mathcal{P}_S$ includes just a single
channel $P_{s_S}$, the compound setting is not required for this
set. For the last set the information index set of the polar coding
construction (in Section~\ref{section:polarCodes}) is therefore
applied. The resulting rate of the parallel scheme is improved and
given by~\eqref{equation:lowerBoundParallelNonDegraded}.
\end{IEEEproof}

\begin{remark}[\textbf{Possible order of channels}]
The channel order may be an important parameter for the provided
parallel scheme (in terms of achievable rates). The channels may be
ordered by their capacity, where
\begin{equation*}
C(P_{s_1}) \leq C(P_{s_2}) \leq \cdots \leq C(P_{s_S}).
\end{equation*}
However, we have no evidence that this order results in the maximal achievable rate (or that it is optimal in any other sense).
\end{remark}

\begin{remark}[\textbf{An upper bound on parallel polar capacity}]
For each set $\mathcal{P}_s$, $s\in[S]$, the upper bound
in~\cite{CompoundCapacity} on the compound capacity can be applied
to upper bound the size of the existing index sets
$\mathcal{A}_n\bigl(\mathcal{P}_s\bigr)$. According
to~\cite[Theorem~5]{CompoundCapacity}, the resulting rate is upper
bounded by
\begin{equation*}
\frac{1}{2^k} \sum_{\mathbf{\sigma}\in\{0,1\}^k} \min_{i\in\{s,\ldots,S\}} I\bigl(P_{s_i}^{\sigma}\bigr)
\end{equation*}
for every $k\in\mathbb{N}$, where $I\bigl(P_{s_i}^{\sigma}\bigr)$
is the capacity of the corresponding tree-channel
$P_{s_i}^{\sigma}$. As mentioned in
Remark~\ref{remark:onTherivationOfCompundBounds}, the actual
derivation in~\cite{CompoundCapacity} is provided for two channels
$P$ and $Q$. Nevertheless, the arguments
in~\cite{CompoundCapacity} are suitable for the case of $S>2$
channels. The proof of the considered upper bound is based on two
major arguments. The first argument is a transformation of a
channel to a sequence of sets of tree-channels (the same as in the
lower bound). Then, for each such set, the maximal achievable rate
is upper bounded by the minimal capacity of the channel
capacities. Since for the last channel set, which is a set of a
single channel, we have no compound setting (as explained in the
proof of Corollary~\ref{corollary:ImprovedPolarCodingScheme}) the
maximal rate at which the parallel polar coding scheme proposed in
Section~\ref{section:TheProposedScheme} can operate reliably is
given by
\begin{equation*}
C(P_{s_S}) + \frac{1}{2^k}  \sum_{s\in[S-1]} \sum_{\mathbf{\sigma}\in\{0,1\}^k} \min_{i\in\{s,\ldots S\}} I\bigl(P_{s_{i}}^{\mathbf{\sigma}}\bigr).
\end{equation*}
An example is provided in~\cite{CompoundCapacity}, demonstrating
that the concerned bound can be smaller than each of the channel
capacities. Specifically, the example in~\cite{CompoundCapacity} is
based on a BSC with a crossover probability of 0.11002 and a BEC
whose erasure probability is 0.5. Both of these channels corresponds
to a capacity of 0.5 bits per channel use. However, as demonstrated
in~\cite[Example~6]{CompoundCapacity}, their compound capacity is
upper bounded by 0.482 bits per channel use. Consequently, if the
parallel polar coding scheme in
Section~\ref{section:TheProposedScheme} is applied for the same two
channels, the possible rate of such a parallel coding scheme is
upper bounded by 0.982 bits per channel use where the parallel
capacity is given by 1 bit per channel use.
\end{remark}

\subsection{Two capacity achieving schemes for non-degraded channels}

Consider the case where transmission takes place over a set of $S$
binary-input, memoryless, output-symmetric channels
$\{P_{s}\}_{s=1}^{S}$. Since the channels are no longer degraded,
the monotonicity property guaranteed in
Corollary~\ref{corollary:informationSetsDegradedChannels} does no
longer apply. Nevertheless, polarization of each one of the
channels is still guaranteed. That is, information set sequences
$\mathcal{A}_n^{(s)}$, $s \in [S]$, satisfying the rate and
performance properties in~\eqref{equation:rateProperties}
and~\eqref{equation:SplitChannelPerformanceProperty} continue to
exists. Two capacity achieving schemes are provided in this
section. The first is based on interleaved binary polar codes, and
the second is based on non-binary polarization.

As in Section~\ref{section:TheProposedScheme}, MDS codes are used
in the parallel coding scheme. Fix an integer $m>0$ such that
$2^m-1\geq S$. All MDS codes to be applied in the introduced
coding scheme are defined over GF$(2^m)$. We assume in the
following that such MDS codes of block length $S$ over GF$(2^m)$
are fixed and known both to the receiver and the transmitter, for
every dimension $d\in [S]$. These MDS codes are denoted by
$\mathcal{C}_{d}$.

\subsection*{Interleaved parallel polar coding scheme}

For the interleaved parallel polar coding scheme, $m$ interleaved
polar codes are applied for every channel $P_{s}$, $s\in [S]$. The
$m$ interleaved polar code of each channel $P_{s}$, $s\in [S]$,
are defined based on the same information set sequence
$\mathcal{A}_n^{(s)}$. The encoding process is defined as follows:
\begin{enumerate}
\item For every information index $k\in \mathcal{A}_n^{(s)}$, and every channel index $s\in [S]$:
\begin{enumerate}
\item Pick $m$ information bits, denoted by $u^{(s)}_{k(m-1)+l}$, $1\leq l\leq m$.
\item Define a symbol $a_s^{(k)}$ over GF$(2^m)$, based on the binary length-$m$ vector
\begin{equation*}
\bigl(u^{(s)}_{(k-1)m+1},\ldots,u^{(s)}_{(k-1)m+m}\bigr).
\end{equation*}
\item For every $k\in [n]$, a length $S$ codeword
$\mathbf{c}^{(k)} = (c_1^{(k)},c_2^{(k)},\ldots,c_S^{(k)})$
over GF$(2^m)$ is defined according to:
\begin{enumerate}
\item Set $d \triangleq |\{s:\ k\in A_n^{(s)}\}|$.
\item Choose the codeword $\mathbf{c}^{(k)} \in \mathcal{C}_d$, satisfying
$ c_{s'}^{(k)} = a_{s'}^{(k)}$ for every $s'\in \{s:\ k\in
A_n^{(s)}\}$. Note that as $\mathcal{C}_d$ is an MDS code of
dimension $d$, the codeword $\mathbf{c}^{(k)}$  is indeed
completely determined by the $d$ indices $\{s:\ k\in A_n^{(s)}\}$.
\end{enumerate}
\end{enumerate}
\item For every index $k\not\in \mathcal{A}_n^{(s)}$ and every $s\in [S]$, define the binary vector
\begin{equation*}
\Bigl(u^{(s)}_{(k-1)m+1},u^{(s)}_{(k-1)m+2},\ldots, u^{(s)}_{(k-1)m+m}\Bigr) \in \{0,1\}^m
\end{equation*}
as the binary vector representation of the symbol $c_{s}^{(k)}$.
\item Compute the $m \cdot S$ polar codewords $\mathbf{x}_{l,s}\in\{0,1\}^n$, $l \in [m]$, $s\in [S]$ according to
\begin{equation*}
\mathbf{x}_{l,s} = \bigl(u^{(s)}_l,u^{(s)}_{m+l},u^{(s)}_{(n-1)m+l}\bigr) \cdot G_n
\end{equation*}
where $G_n$ is the polar generator matrix.
\item For every channel index $s\in [S]$, form a codeword $\mathbf{x}^{(s)}$
for transmission based on the concatenation
\begin{equation*}
\mathbf{x}^{(s)} = (\mathbf{x}_{1,s},\mathbf{x}_{2,s},\ldots,\mathbf{x}_{m,s}).
\end{equation*}
\end{enumerate}

At the decoder, it is assumed that the concatenated codeword
$\mathbf{x}^{\pi(s)}$ is transmitted over the channel $P_{s}$,
$s\in [S]$. The first stage of the decoding process goes as
follows:
\begin{enumerate}
\item For every $s\in [S]$ such that $1 \in A_{n}^{(s)}$, the bits
$u^{(\pi(s))}_{l}$, $l\in [m]$ can be decoded, based on the first
step of the standard polar coding successive cancellation decoding
procedure, for the $m$ interleaved polar codes of the channel $s$.
\item Set $d \triangleq |\{s:\ 1\in A_n^{(s)}\}|$.
\item Find the codeword $\mathbf{c} = (c_1,c_2,\ldots,c_S)$ in
$\mathcal{C}_d$ such that for every $s' \in \{s:\ 1\in
A_n^{(s)}\}$, the symbol $c_{\pi(s')}$ equals to the symbol in
GF$(2^m)$ corresponding to the binary vector
\begin{equation*}
\bigl(u^{(\pi(s'))}_{1},u^{(\pi(s'))}_{2},\ldots,u^{(\pi(s'))}_{m}\bigr)\ .
\end{equation*}
Note that the codeword $\mathbf{c}$ is completely determined by
every $d$ symbols. That is the decoding result does not depend on
the actual permutation $\pi$, applied during the block
transmission.
\item For every $s' \not\in \{s:\ 1\in A_n^{(s)}\}$, the bits
\begin{equation*}
\bigl(u^{(\pi(s'))}_{1},u^{(\pi(s'))}_{2},\ldots,u^{(\pi(s'))}_{m}\bigr)
\end{equation*}
are set to be the length-$m$ binary vector representation of the
symbol $c_{\pi(s')} \in \text{GF} (2^m)$.
\end{enumerate}

Note, that after the first stage of the decoding process, all the
bits $u^{(s)}_l$, $s \in [S]$, $l\in [m]$ are decoded. Next, the
$k$-th stage, $2\leq k\leq n$, of the decoding process is
described.  It is assumed that the decoding of the bits
$u^{(s)}_{m(k'-1)+l}$, $s \in [S]$, $l\in [m]$ are decoded up to
$k'\leq k-1$. The decoding of the bits $u^{(s)}_{m(k-1)+l}$, $s
\in [S]$, $l\in [m]$ goes as follows:
\begin{enumerate}
\item For every $s\in [S]$ such that $k \in A_{n}^{(s)}$, the bits
$u^{(\pi(s))}_{(k-1)m+l}$, $l\in [m]$ can be decoded using $m$
standard polar coding successive cancellation decoding procedures.
These decoding procedures are based on the bits which were decoded
in earlier decoding stage. That is, for a fixed $l$, $l \in [m]$,
the bit $u^{(\pi(s))}_{(k-1)m+l}$ is decoded based on the bits
$u^{(\pi(s))}_{(k'-1)m+l}$, $k'\in [k-1]$ using the standard polar
coding successive cancellation decoding procedure for the polar
code defined based on the index set sequence $A_{n}^{(s)}$.
\item Set $d \triangleq |\{s:\ k\in A_n^{(s)}\}|$.
\item Find the codeword $\mathbf{c} = (c_1,c_2,\ldots,c_S)$ in $\mathcal{C}_d$
such that for every $s' \in \{s:\ k\in A_n^{(s)}\}$, the symbol
$c_{\pi(s')}$ equals to the symbol in GF$(2^m)$ corresponding to
the length-$m$ binary vector representation
\begin{equation*}
\bigl(u^{(\pi(s'))}_{(k-1)m+1},u^{(\pi(s'))}_{(k-1)m+2},\ldots,u^{(\pi(s'))}_{km}\bigr)\ .
\end{equation*}
Note that this codeword is completely determined by every $d$ symbols. That is the decoding result does not depend on the actual permutation $\pi$, applied during the block transmission.
\item For every $s' \not\in \{s:\ k\in A_n^{(s)}\}$, the bits
\begin{equation*}
\bigl(u^{(\pi(s'))}_{(k-1)m+1},u^{(\pi(s'))}_{(k-1)m+2},\ldots,u^{(\pi(s'))}_{km}\bigr)
\end{equation*}
 are set to be the binary vector representation of the symbol $c_{\pi(s')} \in \text{GF}(2^m)$.
\end{enumerate}

\begin{proposition} \label{proposition:CapacityAchievingPropertyInterleavedPolarCodes}
{\em The provided interleaved parallel polar coding scheme
achieves the parallel channel capacity.}
\end{proposition}
\begin{IEEEproof}
Since MDS codes of dimension $d$ posses the property that every set of $d$ symbols completely described a codeword, the performance of the provided decoding process does not depent on the actual transmission permutation. The fact that the resulting error probability approached zero is a direct consequence of the error performance of the channel polarization method. It remains to show that the coding rate approaches capacity. Note that for every channel, $m$ interleaved polar codes of block length $n$ are applied. Hence, for a fixed $n$ the transmission rate is given according to:
\begin{equation*}
\sum_{s=1}^{S} \frac{m \cdot A_n^{(s)}}{m \cdot n}
\end{equation*}
which, according to the polarization propertied in~\eqref{equation:rateProperties}, approaches $\sum_{s=1}^{S} C_s$ as $n$ approached infinitely.
\end{IEEEproof}

\subsection*{Parallel polar coding scheme based on non-binary channel polarization}

As an alternative to $m$ interleaved binary polar codes for every
channel, a single non-binary polar code can be applied. Non-binary
polar code are studied in~\cite{MoriTanakNonBinaryPolarization}
and~\cite{NonBinaryPolar}. For the particular case were the size
of the channel input alphabet is a power of a prime, an explicit
construction is provided in~\cite{MoriTanakNonBinaryPolarization}
in terms of an $n\times n$ generator polarization matrix $G_n$
over GF$(2^m)$. As in the binary polarization method, non-binary
polarization generates information-index set-sequence, for which
the corresponding split channels approaches the perfect channels.
These split channels allow for a corresponding polar successive
cancellation decoding process, while keeping the fraction of
information indices arbitrarily close to the channel capacity.

In order to apply the non-binary polarization coding scheme, a new
set of parallel channels $\{W_s\}_{s=1}^{S}$ is defined according
to
\begin{equation*}
W_s(\mathbf{y}|x) \triangleq \prod_{i=1}^m P_{s}(y_{i}|b_{i})
\end{equation*}
where $\mathbf{y} = (y_1,\ldots,y_m) \in \mathcal{Y}_s$, $x\in\text{GF}(2^m)$, $s \in [S]$, and
\begin{equation*}
\mathbf{b}(x) = \bigl(b_1(x),\ldots,b_m(x)\bigr)
\end{equation*}
is the binary $m$-length vector representation of the symbol $x$.
A coding scheme for the parallel channels $W_s$, $s\in [S]$ is
equivalent to a coding scheme for the original binary parallel
channels where the transmission of a symbol $x$ over a channel
$W_s$ is replaced with $m$ transmissions over the channel $P_s$,
$s\in [S]$. With some abuse of notations, the information index
set sequence for each of the non-binary channels $W_s$, $s\in [S]$
is also denoted by $\mathcal{A}_n^{(s)}$. The encoding for the
parallel non-binary polarization scheme follows according to the
following steps:
\begin{enumerate}
\item For every information index $k\in \mathcal{A}_n^{(s)}$, and
every channel index $s\in [S]$:
\begin{enumerate}
\item Pick $m$ information bits.
\item Denote by $a_s^{(k)}$ the symbol in GF$(2^m)$ corresponding
to these $m$ information bits.
\end{enumerate}
\item For every $k\in [n]$, a length $S$ codeword $\mathbf{c}^{(k)}
= (c_1^{(k)},c_2^{(k)},\ldots,c_S^{(k)})$ over GF$(2^m)$ is defined according to:
\begin{enumerate}
\item Set $d \triangleq |\{s:\ k\in A_n^{(s)}\}|$.
\item Choose the codeword $\mathbf{c}^{(k)} \in \mathcal{C}_d$, satisfying
$c_{s'}^{(k)} = a_{s'}^{(k)}$ for every $s'\in \{s:\ k\in
A_n^{(s)}\}$.
\end{enumerate}
\item Compute $S$ polar codewords $\mathbf{x}_{s}$, $s\in [S]$ according to
\begin{equation}
\mathbf{x}_{s} = \bigl(c_s^{(1)}, c_s^{(2)},\ldots,
c_s^{(n)}\bigr) \cdot G_n
\label{equation:nonBinaryPolarizationCodeWord}
\end{equation}
where $G_n$ is the polar generator matrix and arithmetic is carried over GF$(2^m)$.
\end{enumerate}

The first stage of the decoding process is carried as follows:
\begin{enumerate}
\item For every $s\in [S]$ such that $1 \in A_{n}^{(s)}$, the
symbols $c_{\pi(s)}^{(1)}$ can be decoded, based on the first step
of the polar coding successive cancellation decoding procedure
applied for corresponding non-binary channels $W_{s}$.
\item Set $d \triangleq |\{s:\ 1\in A_n^{(s)}\}|$.
\item Find the codeword $\mathbf{c} = (c_1,c_2,\ldots,c_S)$ in
$\mathcal{C}_d$ such that for every $s' \in \{s:\ 1\in
A_n^{(s)}\}$
\begin{equation} \label{equation:MDSCodeWordConditionNonBinaryPolarization}
c_{\pi(s')}^{(1)} = c_{\pi(s')}.
\end{equation}
\item For every $s' \not\in \{s:\ 1\in A_n^{(s)}\}$, decode the
symbols $c_{\pi(s)}^{(1)}$ according to~\eqref{equation:MDSCodeWordConditionNonBinaryPolarization}.
\end{enumerate}
Next, assume that the decoding process is complete up to step
$k-1$ where $2\leq k \leq n$. That is, for every $k' \in [k-1]$
the symbols $c_{s}^{(k')}$, $s \in [S]$ are already decoded. The
decoding of the symbols $c_{s}^{(k)}$, $s \in [S]$, is carried as
follows:
\begin{enumerate}
\item For every $s\in [S]$ such that $k \in A_{n}^{(s)}$, the symbols $c_{\pi(s)}^{(k)}$
can be decoded based on the channel observations and the former
decoded symbols $c_{s}^{(k')}$, $s\in [S]$ and $k' \in [k-1]$. The
symbol $c_{\pi(s)}^{(k)}$ is decoded using the polar coding
successive cancellation decoding procedure applied for
corresponding non-binary channels $W_{s}$ and depends on the
former decoded symbols $c_{\pi(s)}^{(k')}$, $k' \in [k-1]$.
\item Set $d \triangleq |\{s:\ k\in A_n^{(s)}\}|$.
\item Find the codeword $\mathbf{c} = (c_1,c_2,\ldots,c_S)$ in
$\mathcal{C}_d$ such that for every $s' \in \{s:\ k\in A_n^{(s)}\}$
\begin{equation} \label{equation:MDSCodeWordCondition2NonBinaryPolarization}
c_{\pi(s')}^{(k)} = c_{\pi(s')} .
\end{equation}
\item For every $s' \not\in \{s:\ k\in A_n^{(s)}\}$, decode
the symbols $c_{\pi(s)}^{(k)}$ according
to~\eqref{equation:MDSCodeWordCondition2NonBinaryPolarization}.
\end{enumerate}

The same reasoning as in
Proposition~\ref{proposition:CapacityAchievingPropertyInterleavedPolarCodes}
shows that the non-binary scheme also achieves the capacity for
the provided model. What is left to provide is a generalization of
the symmetry results in~\cite{ArikanPolarCodes} discussed in
Remark~\ref{remark:WhyWeNeedSymmetry}. Specifically, it remains to
show that the the error performance of the non-binary polar codes
under successive cancellation is independent on the symbol vectors
$\bigl(c_s^{(1)}, c_s^{(2)},\ldots, c_s^{(n)}\bigr)$, $s\in[S]$,
in~\eqref{equation:nonBinaryPolarizationCodeWord}. This result is
provided in the Appendix and follows along similar arguments as
in~\cite{ArikanPolarCodes}.

\section{Summery and Conclusions} \label{section:conclusions}

Parallel polar coding schemes are provided in this paper for
communicating over a set of parallel binary-input memoryless and
output-symmetric parallel channels. The provided coding schemes
are based on the channel polarization method originally introduced
by Arikan~\cite{ArikanPolarCodes} for a single-channel setting.
The first provided scheme is shown to achieve capacity for the
particular case of stochastically degraded channels. For
non-degraded parallel channels, upper and lower bounds on the
achievable rates are derived for the provided scheme based on the
techniques in~\cite{CompoundCapacity}. Two modifications of the
parallel polar coding scheme are provided, which achieve the
capacity of the general non-degraded case.

The definition of polar codes includes a set of predetermined and
fixed bits. These bits are crucial to the decoding process. In the
original polarization scheme in~\cite{ArikanPolarCodes}, these
predetermined and fixed bits may be chosen arbitrarily (in the
case of symmetric channels). For the provided parallel coding
schemes on the other hand, the predetermined and fixed bits are
determined based on algebraic coding constraints. For the
particular case of degraded channels, the information bits of
channels determine the predetermined and fixed bits of their
degraded counterparts. The MDS coding, suggested in this paper is
similar to the rate-matching scheme
in~\cite{WillemsGorolhovParallelChannels}. For the general
non-degraded case, either interleaving of binary polar codes is
used or non-binary channel polarization. The modification based on
non-binary channel polarization is almost directly applicable for
the case of non-binary parallel channels.

The following topics are considered for further research:
\begin{enumerate}
\item Symmetry condition: For symmetric channels, the predetermined
and fixed bits may be chosen arbitrarily. For non-symmetric
channels, good predetermined and fixed bits (called also frozen
bits in~\cite{ArikanPolarCodes}) are shown to exist, but their
choice may not be arbitrary. It is an open question if there is a
more general construction that does not require symmetry of the
parallel channels. This may be accomplished using non-binary codes
(the single channel case is addressed in some extent
in~\cite{SatishPHD}).
\item Generalized parallel polar coding as in
\cite{Satish1}-\cite{Arikan2D}.
\item Generalized channel models. Arbitrarily-permuted channels is
just one particularization of the compound setting. It is of great
interest to enlarge the family of parallel channels for which the
studied coding scheme may apply. Of specific interest is the case
of parallel channels were a sum-rate constraint is provided by the
channel model characterization.
\end{enumerate}

\section*{Acknowledgment}

This research was supported by the Israel Science Foundation (grant
no. 1070/07), and by the European Commission in the framework of the
FP7 Network of Excellence in Wireless Communications (NEWCOM++).


\appendix

In this appendix we show that the performance of non-binary
polarization provided in~\cite{MoriTanakNonBinaryPolarization}
and~\cite{NonBinaryPolar} under successive cancellation decoding
is independent on the input vectors (which includes both
information and predetermined and fixed symbols). The applied
proof techniques goes along a similar steps as in the binary case
provided in~\cite{ArikanPolarCodes}. We consider a polarization
scheme where transmission takes place over a DMC whose input
alphabet $\mathcal{X}$ is a finite field. It is assume that the
polarization scheme can be defined according
to~\eqref{equation:equivalentRecursiveConstruction}, where all
operations are carried over the considered finite field. in order
to achieve message independence property, we relay on the
following symmetry definition for the non-binary case:
\begin{definition}[\textbf{Non-binary symmetry}] \label{definition:nonBinarySymmetry}
A memoryless channel which is characterized by a transition
probability $p$, an input-output alphabet $\mathcal{X}$ and a
discrete output alphabet $\mathcal{Y}$ is {\em symmetric} if there
exists a function $\mathcal{T}:\ \mathcal{Y}\times \mathcal{X} \to
\mathcal{Y}$ which satisfies the following properties:
\begin{enumerate}
\item For every $x\in\mathcal{X}$, the function $\mathcal{T}(\cdot,x):\ \mathcal{Y} \to \mathcal{Y}$ is bijective.
\item For every $x_1, x_2 \in \mathcal{X}$ and $y\in\mathcal{Y}$, the following equality holds:
\begin{equation*}
p\bigl(y|x_1\bigr) = p\bigl(\mathcal{T}(y,x_2-x_1)|x_2\bigr).
\end{equation*}
\end{enumerate}
\end{definition}

Let $p$ be a symmetric DMC with an input alphabet $\mathcal{X}$ and output alphabet $\mathcal{Y}$. In addition, let $\mathcal{T}$ be the corresponding function in Definition~\ref{definition:nonBinarySymmetry}. With abuse of notation, the operation of $\mathcal{T}$ on vectors $\mathbf{y}\in\mathcal{Y}^n$ and $\mathbf{x}\in\mathcal{X}^n$ is carried according to
\begin{equation*}
\mathcal{T}(\mathbf{y}, \mathbf{x}) \triangleq \bigl( \mathcal{T}(y_1,x_1), \ \mathcal{T}(y_2,x_2),\ \ldots, \mathcal{T}(y_n,x_n)\bigr).
\end{equation*}
Subtraction of a vector is also defined item-wise, that is $-(x_1,\ldots,x_n) = (-x_1,\ldots x_n)$.

The polar successive cancellation decoding is accomplished based
on decision made according to the split channel output
probabilities. For the case of non-binary polarization, the
corresponding split channels are defined according to
\begin{equation} \label{equation:nonBinaryPolarChannelSplitting}
p_n^{(l)}(\mathbf{y},\mathbf{w}|x) \triangleq
\frac{1}{|\mathcal{X}|^{n-1}}\sum_{\mathbf{c} \in \mathcal{X}^{n-l}}
p_n\bigl(\mathbf{y}|(\mathbf{w},x,\mathbf{c})\bigr),\quad l\in[n]
\end{equation}
where $\mathbf{y} \in \mathcal{Y}^n$, $\mathbf{w} \in
\mathcal{X}^{l-1}$, and $x \in \mathcal{X}$. Note that this
definition transforms to the binary base
in~\eqref{equation:standradPolarChannelSplitting} for binary input
alphabets. The error event under successive cancellation is a
subset of the following union
\begin{equation*}
\bigcup_{d\in\mathcal{X}\setminus\{0\}}\bigcup_{i\in\mathcal{A}} \mathcal{E}_i^d
\end{equation*}
where $\mathcal{A}$ is the set of indices of split channels which polarizes to perfect channels and
\begin{equation} \label{equation:specificErrorEventNonBinaryPolarization}
\mathcal{E}_i^d \triangleq \Bigl\{ (\mathbf{w}, \mathbf{y}) \in \mathcal{X}^n \times \mathcal{Y}^n:\ p_n^{(i)}\bigl(\mathbf{y},(w_1,\ldots,w_{i-1})|w_i\bigr) \leq p_n^{(i)}\bigl(\mathbf{y},(w_1,\ldots,w_{i-1})|w_i+d\bigr) \Bigr\}.
\end{equation}
On the other hand, non-binary channel polarization guarantees that
there a symbol set $\{w_i\}_{i\in [n]\setminus \mathcal{A}}$ such
that the probability of the event $\mathcal{E}_i^d$ approaches
zero for every $d\in\mathcal{X}\setminus\{0\}$ and
$i\in\mathcal{A}$.

The following lemma assures that for symmetric channels the events
$\mathcal{E}_i^d$, $i\in\mathcal{A}$ and
$d\in\mathcal{X}\setminus\{0\}$, are independent with the input
vector $\mathbf{w}$
in~\eqref{equation:equivalentRecursiveConstruction}. Consequently,
for symmetric channels the error performance guaranteed by
non-binary channel polarization
in~\cite{MoriTanakNonBinaryPolarization} is provided no matter
what are the symbols chosen for $\{w_i\}_{i\in [n]\setminus
\mathcal{A}}$.

\begin{lemma}[\textbf{Message independence property for non-binary symmetric-channel polarization}]
{\em Denote by $P_{\text{e}}(\mathcal{E}_i^d | \mathbf{u})$ the
probability of the event $\mathcal{E}_i^d$
in~\eqref{equation:specificErrorEventNonBinaryPolarization},
assuming that $\mathbf{w} = \mathbf{u}$
in~\eqref{equation:equivalentRecursiveConstruction}. Then,
\begin{equation*}
P_{\text{e}}(\mathcal{E}_i^d | \mathbf{u}) = P_{\text{e}}(\mathcal{E}_i^d | \mathbf{0})
\end{equation*}
for every $\mathbf{u} \ in \mathcal{X}^n$, where $\mathbf{0}$ is the all zero vector in $\mathcal{X}^n$.}
\end{lemma}

\begin{IEEEproof}
Based on the symmetry property of that channel, for every $i\in [n]$, $\mathbf{y}\in\mathcal{Y}^n$, $\mathbf{w}\in\mathcal{X}^{i-1}$, $w\in\mathcal{X}$ and $\mathbf{a}\in\mathcal{X}^n$, we have
\begin{align*}
& p_n^{(i)}\bigl(\mathbf{y},(w_1,\ldots,w_{i-1})|w_i\bigr) \\
& \stackrel{\text{(a)}} = \frac{1}{|\mathcal{X}|^{n-1}}\sum_{\mathbf{c} \in \mathcal{X}^{n-l}}
\prod_{t=1}^n p\Bigl(y_t|\bigl((w_1,\ldots,w_{i-1},x,\mathbf{c})G_n\bigr)_t\Bigr)\\\
\label{equation:symmetryPropertyStepForSplitChannels}
& \stackrel{\text{(b)}} = \frac{1}{|\mathcal{X}|^{n-1}}\sum_{\mathbf{c} \in \mathcal{X}^{n-l}}
\prod_{t=1}^n p\Biggl(\mathcal{T}\Bigl(y_t, \bigl(\mathbf{a}G_n\bigr)_t\Bigr) |\bigl((\mathbf{w},x,\mathbf{c})G_n\bigr)_t + \bigl(\mathbf{a}G_n\bigr)_t \Biggr)
\end{align*}
where $(\mathbf{x})_t$ denotes the $t$-th element of a vector
$\mathbf{x}=(x_1,\ldots,x_n)$, (a) follows for memoryless channels
from~\eqref{equation:equivalentRecursiveConstruction}
and~\eqref{equation:nonBinaryPolarChannelSplitting} and (b)
follows from the symmetry property of the channel. Consequently,
it follows that
\begin{equation} \label{equation:symmetryPropoertyForNonBinarySplitChannels}
p_n^{(i)}\bigl(\mathbf{y},(w_1,\ldots,w_{i-1})|w_i\bigr) = p_n^{(i)}\Bigl(\mathcal{T}\bigl(\mathbf{y}, \mathbf{a}G_n, (w_1,\ldots,w_{i-1}) + (a_1,\ldots,a_{i-1})\bigr)|w_i + a_i\Bigr).
\end{equation}
From,~\eqref{equation:specificErrorEventNonBinaryPolarization}
and~\eqref{equation:symmetryPropoertyForNonBinarySplitChannels},
it follows for every pair $(\mathbf{w},
\mathbf{y})\in\mathcal{X}^n \times \mathcal{Y}^n$ and every
$\mathbf{a}\in\mathcal{X}^n$ that
\begin{equation} \label{equation:technicalPropertyOfErrorEvent}
\bigl(\mathbf{w},\mathbf{y}\bigr) \in \mathcal{E}_i^d \iff \bigl(\mathbf{a}+\mathbf{w}, \mathcal{T}(\mathbf{y},\mathbf{a}\cdot G_n)\bigr) \in \mathcal{E}_i^d.
\end{equation}
Next, let $1_{\mathcal{E}_i^d}(\mathbf{u},\mathbf{y})$ denote the
indicator of the event $\mathcal{E}_i^d$. For every
$\mathbf{u}\in\mathcal{X}^n$ it follows that
\begin{align*}
& P_{\text{e}}(\mathcal{E}_i^d | \mathbf{u}) \\
& = \sum_{\mathbf{y} \in\mathcal{Y}^n} p_n(\mathbf{y}|\mathbf{u})1_{\mathcal{E}_i^d}(\mathbf{u},\mathbf{y}) \\
& \stackrel{\text{(a)}} = \sum_{\mathbf{y} \in\mathcal{Y}^n} p(\mathbf{y}|\mathbf{u} G_n) 1_{\mathcal{E}_i^d}(\mathbf{u},\mathbf{y}) \\
& \stackrel{\text{(b)}}= \sum_{\mathbf{y} \in\mathcal{Y}^n} p(\mathcal{T}(\mathbf{y},-\mathbf{u}G_n) | \mathbf{0}) 1_{\mathcal{E}_i^d}(\mathbf{0},\mathcal{T}(\mathbf{y},-\mathbf{u}G_n)) \\
& = \sum_{\mathbf{y} \in\mathcal{Y}^n} p_n(\mathbf{y}|\mathbf{0})1_{\mathcal{E}_i^d}(\mathbf{0},\mathbf{y})\\
& = P_{\text{e}}(\mathcal{E}_i^d | \mathbf{0})
\end{align*}
where (a) follows
from~\eqref{equation:equivalentRecursiveConstruction}, (b) follows
from~\eqref{equation:technicalPropertyOfErrorEvent} by plugging
$\mathbf{a}=\mathbf{u}$, and (c) follows since $\mathcal{T}(y,x)$
is a bijective function of $y\in\mathcal{Y}$ for every fixed
symbol $x\in\mathcal{X}$.
\end{IEEEproof}

\end{document}